\documentclass[10pt]{book}

\usepackage{sussp2004}
\usepackage{epsfig}
\usepackage{graphicx}
\usepackage{psfrag}
\usepackage{latexsym}

\begin{document}

\setcounter{section}{1}
\setcounter{page}{1}

\headings{Lattice QCD - A guide for people who want results}
{Lattice QCD}
{Christine Davies }
{Department of Physics and Astronomy,\\University of Glasgow, UK}

\section{Introduction}
Lattice QCD \index{lattice QCD}was invented thirty years ago but 
only in the last few years has it finally fulfilled its promise 
as a precision tool for calculations in hadron physics. 
This review will cover the fundamentals of discretising QCD onto 
a space-time lattice and 
how to reduce the 
errors associated with the discretisation. This `improvement' is 
the key that has made the enormous computational task of a lattice 
QCD calculation tractable and enabled 
us to reach the recent milestone of precision calculations of simple 
`gold-plated' hadron masses. 
Accurate decay matrix elements, such as those for leptonic 
and semileptonic decay of heavy mesons needed by the $B$ factory 
experimental programme, are now within sight. I will describe what goes 
into such calculations and what the future prospects and limitations are. 

\section{Lattice QCD formalism and methods}

\subsection{The Path Integral}

Lattice QCD is based on the path integral formalism. We
can demonstrate this formalism by discussing the 
solution of the quantum mechanical problem of a 
particle moving in one dimension (Lepage 1998a). 
This could be solved using Schr\"{o}dinger's equation with 
$H=p^2/2m+V(x)$ and $[x,p]=i$ (in
`particle physics units', $\hbar=c=1$). We can 
also solve it using the path integral formulation
and this is the basis for lattice QCD. 

\begin{figure}[ht]
\centering{
\includegraphics[width=80mm, clip, trim=0 0 0 0]{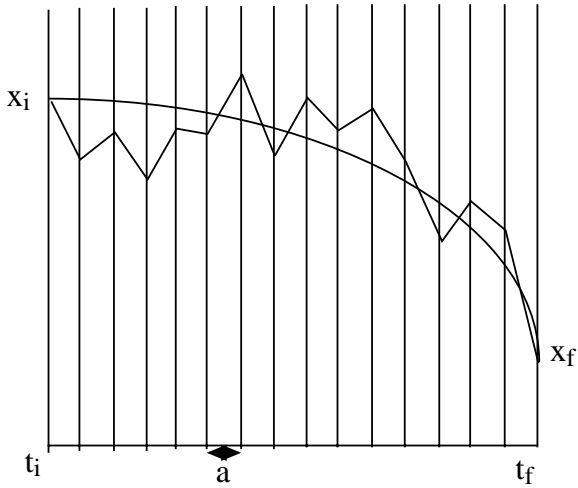}
\caption{Discretised classical and possible quantum 
mechanical paths from $x_i$ to $x_f$ for a particle 
moving in one dimension. }
\label{davies:fig01}}
\end{figure}

A key quantity is the transition amplitude between 
eigenstates of position at, say, time $t=t_i$ and $t_f$.
This is expressed as a functional integral over 
all possible paths $x(t)$ from $t=t_i$ to $t_f$ weighted
by the exponential of the action, $S$, which is the integral of 
the Lagrangian, $L$. 
\begin{eqnarray}
\langle x_f(t_f) |  x_i(t_i) \rangle = \int {\cal{D}}x(t)e^{iS[x]} \label{pathintM} \\
S[x] \equiv \int_{t_i}^{t_f} dt L(x,\dot x) \equiv \int dt [\frac {m \dot x(t)^2}{2} - V(x(t))].
\label{SintM}
\end{eqnarray}
The path integral can be evaluated by discretising 
time into a set of points, $t_j = t_i +ja$ for $j = 0,1 \ldots N$ 
and $a$ the lattice spacing $\equiv (t_f-t_i)/N$. The
path $x(t)$ then becomes a set of variables, $x_j$, 
and Equation~\ref{pathintM} above 
requires us to integrate over each one, {\textit{ie}} the problem reduces 
to an ordinary integral but over an $N-1$-dimensional space. 
The end points, $x_0$ 
and $x_N$ are kept fixed in this example. The most likely path 
is the classical one, which minimises the action ($m\ddot x = V^{\prime}$), 
but the path integral allows quantum fluctuations about this, 
see Figure~\ref{davies:fig01}.  

The $i$ in front of the action in the exponential gives rise 
to the problem, both conceptual and numerical, of adding 
oscillating quantities together. It is simpler to rotate 
the time axis to Euclidean time, $t \rightarrow -it$. Then 
the path integral becomes 
\begin{eqnarray}
\int {\cal{D}}x(t)e^{-S[x]} = A \int_{-\infty}^{\infty} dx_1dx_2\ldots dx_{N-1} e^{-S[x]} \nonumber \\
S[x] \equiv \int_{t_i}^{t_f}  L(x,\dot x) \equiv \sum_{j=0}^{N-1} [\frac {m \dot x_j^2}{2} + V(x_j)],
\label{pathintE}
\end{eqnarray}
where the integrals over the intermediate points 
in the path, $x(t)$, are now explicit. We will ignore the normalisation of the 
integral, $A$. 
To perform the integral we must discretise the Lagrangian so 
that it takes values at the discrete time points. Then 
\begin{equation}
S = \sum_{j=0}^{N-1}[\frac{m}{2a}(x_{j+1}-x_{j})^2 +aV(x_j)].
\end{equation}
Clearly the accuracy of our discretisation will depend on 
$a$ being small. However, as $a$ becomes smaller at fixed physical 
time length, the number of points, $N$, increases and so does the 
computational cost. 

For large $N$ an efficient way to perform the integral is 
by Monte Carlo. A set of possible values for $x_j,\, j=1,N$ is 
called a {\it configuration}. The configurations with most 
weight in the integral are those with large $e^{-S}$. For 
maximum efficiency we want to
generate configurations with probability 
$e^{-S}$ - this is known as {\it importance sampling}. 
A simple method for doing this is the Metropolis algorithm. 
This starts with an initial configuration ({\textit{eg}} $x_j$ all zero 
or chosen randomly). It then passes through the $x_j$
in turn proposing a change to a given $x_j$ of size $\epsilon$, 
{\textit{ie}} a random number between $-\epsilon$ 
and $\epsilon$ temporarily added to $x_k$, say. 
The change in the action as a result of 
$x_k$ changing is calculated. Call this $\Delta S$. Note that this 
calculation only involves the $x_j$ in the neighbourhood of $x_k$ 
and connected to it through the action. If $\Delta S < 0$ the change
is accepted. If $\Delta S > 0$ another random number, uniformly 
distributed between 0 and 1 is generated. If $e^{-\Delta S} > rand$ 
then the change is accepted. If not, $x_k$ reverts to its previous 
value. (Lepage 1998a, di Pierro 2000)

In this way a new configuration is generated after each sweep through 
the lattice. A set of configurations is called an {\it ensemble}. 
Calculations of various functions of the $x_j$ on an ensemble 
then yield the physics results that we are after, such as the 
quantised energy levels available to the particle. 

In wavefunction language we can express 
\begin{equation}
\langle x_f | e^{-iH(t_f-t_i)} | x_i \rangle = \sum_n \psi^*_n(x_f)\psi_n(x_i)e^{-iE_n(t_f-t_i)} ,
\end{equation}
by inserting a complete set of eigenstates of the Hamiltonian, $H$.
When rotated to Euclidean time, and taking $t_f-t_i=T$ and $x_i=x_f=x$ we have
\begin{equation}
\langle x | e^{-HT} | x \rangle = \sum_n \psi^*_n(x)\psi_n(x)e^{-E_nT}, 
\end{equation}
where $\psi_n(x) = \langle n | x \rangle$.
It now becomes clear that the result will be dominated by the 
ground state as $T$ becomes large, because all higher states are 
exponentially suppressed. Integrating over the initial and final 
$x$ values gives: 
\begin{equation}
\int dx \langle x | e^{-HT} | x \rangle \rightarrow e^{-E_0T}, T \rightarrow \infty.
\end{equation}
This would be the result, up to a normalisation, of integrating 
by Monte Carlo over all the $x_j$, setting the initial and final 
values to be the same, {\textit{ie}} using periodic boundary conditions. 
The result looks very similar to 
a problem in statistical mechanics when formulated in this way, 
as did Equation~\ref{pathintE}, and this is not an accident. 
Indeed we can work at non-zero temperature if we take $T$ finite, 
but here we will concentrate on the zero temperature case where 
in principle $T \rightarrow \infty$, and in 
practice is large. 

To investigate excitations 
above the ground state, we must interrupt the propagation of 
the ground state by introducing new operators at intermediate 
times. For example,
\begin{eqnarray}
\frac {\langle x(T) | x(t_2) x(t_1)| x(0) \rangle}
{\langle x(T) | x(0) \rangle} &=& \frac {\int {\cal{D}}x\,\, x(t_2) x(t_1) e^{-S[x]}} 
{\int {\cal{D}}x \, e^{-S[x]}} \nonumber \\
&=& |\langle E_0 | x | E_1 \rangle|^2 e^{-(E_1-E_0)(t_2-t_1)}, t_2-t_1 \rightarrow \infty.
\label{x1x2}
\end{eqnarray}
The state propagating between the insertions of the $x$ operators at 
$t_1$ and $t_2$ cannot be the ground state, since $x$ switches parity. 
If $t_2 - t_1$ is large enough then the first excited state will 
dominate, by the same argument as used above for ground state 
domination. So, if we can evaluate the ratio of path integrals, 
we can determine the energy splitting $E_1 - E_0$ between the 
first excited and the ground state. 
The evaluation is very simply done by taking an ensemble of configurations 
generated by the Metropolis algorithm above and `measuring' on each 
one the value of $x(t_2)x(t_1)$. The ensemble average of this 
quantity, denoted $<<x(t_2)x(t_1)>>$, is then the ratio above. 

The evaluation will suffer from a statistical error depending 
on how many configurations are in the ensemble. This will 
improve as the square root of the number of configurations provided that the 
results on each configuration are statistically independent. 
Since the configurations were made in a sequence, this will 
not be strictly true and results on neighbouring configurations 
in the sequence will be correlated. There are various statistical 
techniques, such as binning (averaging) neighbouring results and 
then recalculating the statistical error, that will 
uncover correlations. If the statistical error grows with 
the bin size then the results are correlated and the binned 
results should be used rather than the raw results. It might also be true that results on 
some number of the initial configurations of an ensemble have
to be discarded because these configurations, and measurements 
on them, are not 
yet typical of the $e^{-S}$ distribution. For this example, 
the results can obviously be improved statistically by 
moving $t_1$ and $t_2$ along the time axis, keeping $t_2-t_1$ 
fixed, and averaging those results.  

To extract $E_1 - E_0$ we should fit the results as 
a function of $t_2-t_1$ to the exponential form of Equation~\ref{x1x2}. Since 
$t_2-t_1$ will not be infinite, we can take account of 
contamination from higher states by fitting to a sum of 
exponentials in which we constrain the higher $E_n > E_1$. 
Finally we may also improve the accuracy on the energy 
by modifying the operator $x(t_1)x(t_2)$ to a product of 
functions of $x(t_{1,2})$ which has optimum overlap with 
the first excited state and minimum overlap with higher 
excited states. This will allow Equation~\ref{x1x2} to 
become true for smaller values of $t_2-t_1$ and improve 
the extraction of $E_1-E_0$ from the fits. Indeed we 
can fit simultaneously to the results from several different operators
and this again will improve the accuracy with which $E_1-E_0$ 
can be determined. 
It is a useful exercise to do the calculation above for, 
say, the simple harmonic oscillator potential, $V=x^2/2$ and $m=1$. 
All the pitfalls of `measurement' and the improvements above have their 
mirror in lattice QCD calculations, as we shall see, but
this simple example demonstrates them very clearly in a setting 
where computer time is not an issue. 

An important issue is that of the systematic errors, called 
discretisation errors, introduced because of the non-zero 
lattice spacing. We can make these explicit by 
rewriting the finite difference in terms of the 
continuum ({\textit{ie}} continuous space-time of the real world) 
derivative. The exponential of the continuum 
derivative is the translation operator for moving from one site 
to the next. So
\begin{equation}
\frac{x_{i+1}-x_i}{a} = \frac{e^{\partial a} - 1}{a} = \partial + \frac{a\partial^2}{2}.
\end{equation}
This shows that this form of the finite difference has errors linear in $a$. 
The actual size of the errors at a given value of $a$ would depend on the 
effective size of $\partial^2$ in the quantity being calculated. To halve 
the errors requires half the lattice spacing at (at least) double the 
computer cost. 

The situation is significantly improved by using an improved discretisation.
$\Delta x_i = (x_{i+1}-x_{i-1})/2a$ has errors first at $\cal{O}$$(a^2)$. Further 
improvement is obtained by correcting for the $a^2$ error using 
\begin{equation}
\Delta x_i - \frac{a^2}{6}(\Delta)^3 x_i
\end{equation}
where $(\Delta)^3$ is a discretisation of the third 
order derivative. Errors are then $\cal{O}$$(a^4)$ and this means 
very much smaller errors at a given value of $a$, or a huge 
reduction in computer cost for a given error by being able to 
work at a larger value of $a$. The improved discretisation 
costs a little in computer power to implement but this is 
completely negligible compared to the computer cost saving of the 
improvement and it is this that has made lattice QCD calculations 
tractable. 

\subsection{Gluon fields in lattice QCD}

\begin{figure}[ht]
\centering{
\includegraphics[width=70mm, clip, trim=0 0 0 0]{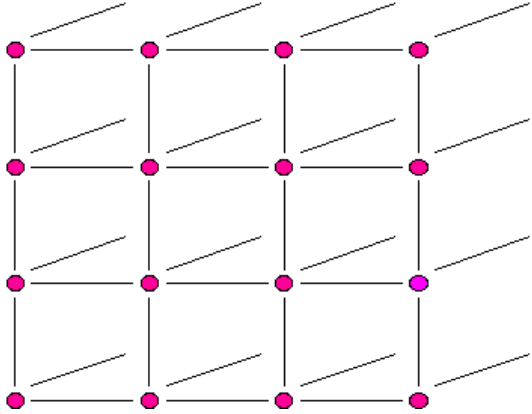}
\caption{A 2-dimensional rendition of a 3-dimensional cubic lattice. Lattice QCD calculations 
use a 4-dimensional grid.}
\label{davies:fig02}}
\end{figure}

Now we have all the ingredients for lattice QCD. 
The position operator as a function of time is 
replaced by the quark and gluon field operators as 
a function of four-dimensional space-time. 
We discretise the space-time into a lattice of 
points (\Figure{\ref{davies:fig02}}) in order to be able to calculate 
the Feynman Path Integral numerically, using Monte 
Carlo methods on a computer. The ground-state which 
dominates the path integral is the QCD vacuum and 
we will be interested in excitations of this which 
correspond to hadrons. 

It is important to remember that lattice QCD is not 
complicated. It is a straightforward simulation 
of the theory on a computer, but a numerically intensive one. 
It is the theory of QCD that is being simulated, not a 
model. However, we are limited in the things that we 
can calculate. There are also statistical and systematic 
errors associated with the calculations and it is important 
to understand where they come from, so that you can 
assess the usefulness of a particular calculation
for your needs. 

Consider the operator 
${\mathcal{O}} = (\overline{\psi}\psi)_y (\overline{\psi}\psi)_x$.
This creates a hadron at a point $x$ and destroys it
at a point $y$. This is the QCD generalisation of 
the $x(t_1)x(t_2)$ operator of the previous section. 
Then the matrix element in the vacuum of the operator 
is given in path integral form as:
\begin{equation}
\langle 0|{\mathcal{O}}|0 \rangle =
\Frac {\int{[d \psi]\,[d \bar{\psi}]\,[d A_{\mu}] \,{\mathcal{O}}[\psi,\overline{\psi},A] e^{-S}}}
{\int{[d \psi]\,[d \bar{\psi}]\,[d A_{\mu}] e^{-S_{QCD}}}}.
\label{fpi}
\end{equation}

The path integral runs over all values of the quark and gluon
fields $\psi$ and $A$ at every point in space-time. 
Discretisation of space-time onto a lattice makes 
the number of space-time points (and therefore field variables) finite.
Continuous space-time $(x,t)$ becomes a grid of
labelled points, $(x_i,t_i)$ or $(n_ia,n_ta)$ where $a$ is the lattice spacing
(this doesn't have to be the same in all directions but usually 
is). 
The fields are then
associated only with the sites, $\psi(x,t) \rightarrow
\psi(n_i,n_t)$. The action must also be discretised, but, as
in the previous section, this is 
straightforward, replacing
fields by fields at the
lattice sites and the derivatives by finite differences of
these fields. The integral over space-time of the Lagrangian becomes a
sum over all lattice sites: ($\int{d^4x} \rightarrow \sum_n a^4$), 
and the path integral becomes a product of integrals over 
each of the fields, to be done by Monte Carlo averaging.
These integrals will be finite, unlike such integrals in 
the continuum, because the lattice provides a regularisation 
of the theory. The lattice spacing provides an ultra-violet 
cut-off in momentum space 
since no momenta larger than $\pi/a$ make sense (since the wavelength is
then smaller than $a$). 

To discretise gauge theories such as QCD
onto a lattice in fact requires a little additional thought to 
that described above because of the
paramount importance of local gauge invariance. The r\^{o}le of the
gluon (gauge) field in QCD is to transport colour from one place to
another so that we can rotate our colour basis locally. It should then
seem natural for the gluon fields to `live' on the links connecting
lattice points, if the quark fields `live' on the sites.

\begin{figure}[ht]
\centering{
\includegraphics[width=90mm, clip, trim=0 35 60 0]{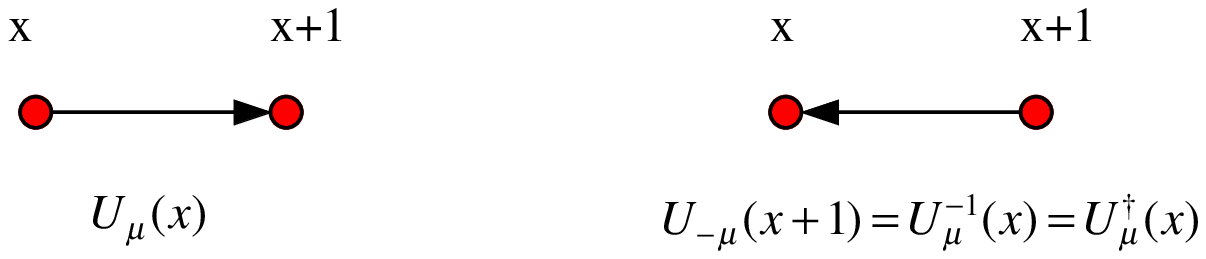}
\vspace{1ex}
{\rule{10mm}{0mm}$U_\mu(x)$\rule{35mm}{0mm}$U_{-\mu}(x+1)=U_\mu^{-1}=U_\mu^\dagger(x)$}\\
\caption{The gluon field on the lattice.}
\label{davies:fig03}}
\end{figure}

The gluon field is also expressed somewhat differently on the lattice
to the continuum. The continuum $A_{\mu}$ is an 8-dimensional vector,
understood as a product of coefficients $A^b_{\mu}$ times the 8
matrices, $T_b$, which are generators of the SU(3) gauge group for
QCD. On the lattice it is more useful to take the gluon field on each
link to be a member of the gauge group itself {\textit{ie}} a special
(determinant = 1) unitary $3\times3$ matrix. The lattice gluon field
is denoted $U_{\mu}(n_i,n_t)$, where $\mu$ denotes the direction of
the link, $n_i,n_t$ refer to the lattice point at the beginning of the
link, and the colour indices are suppressed.  We will often just revert
to continuum notation for space-time, as in $U_{\mu}(x)$.  The lattice
and continuum fields are then related exponentially,
\begin{equation}
U_{\mu} = e^{-iagA_{\mu}} 
\label{ulattcont}
\end{equation}
where the $a$ in the exponent makes it dimensionless, and we include the 
coupling, $g$, for convenience.
If $U_{\mu}(x)$ is the gluon field connecting the points $x$ and 
$x+1_{\mu}$
(see \Figure{\ref{davies:fig03}}), 
then the gluon field connecting these same points but in the 
downwards direction must be the inverse of this matrix, $U_{\mu}^{-1}(x)$.
Since the $U$ fields are unitary matrices, satisfying $U^{\dag}U=1$, 
this is then $U^{\dag}_{\mu}(x)$. 

This form for the gluon field makes it possible to maintain exact
local gauge invariance on a lattice. \index{gauge invariance, lattice
QCD}To apply a gauge transformation to a set of gluon fields we must
specify an SU(3) gauge transformation matrix at each point. Call this
$G(x)$. Then the gluon field $U_{\mu}(x)$ simply gauge transforms by
the (matrix) multiplication of the appropriate $G$ at both ends of its
link. The quark field (a 3-dimensional colour vector) transforms by
multiplication by $G$ at its site.
\begin{eqnarray}
U_{\mu}^{(g)}(x) &=& G(x)U_{\mu}(x)G^{\dag}(x+1_{\mu}) \nonumber \\
\psi^{(g)}(x) &=& G(x)\psi(x) \nonumber \\
\overline{\psi}^{(g)}(x) &=& \overline{\psi}(x)G^{\dag}(x)
\label{gts}.
\end{eqnarray}
To understand how this relates to continuum gauge transformations
try the exercise of setting $G(x)$ to a simple U(1) transformation,
$e^{-i\alpha(x)}$, and show that Equation~\ref{gts} is equivalent to the 
QED-like gauge transformation in the continuum, $A^g_{\mu} =
A_{\mu} - \partial_{\mu}\alpha$.  

\begin{figure}[ht]
\centering{
\includegraphics[width=110mm, clip, trim=0 0 0 0]{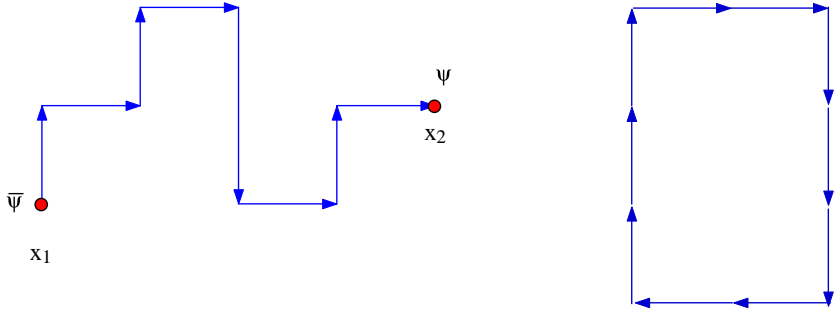}
\caption{A string of gluon fields connecting quark and antiquark
fields (left) and a closed loop of gluon fields (right).}
\label{davies:fig04}}
\end{figure}

Gauge-invariant objects can easily be made on the lattice (see 
\Figure{\ref{davies:fig04}}) out of
closed loops of gluon fields or strings of gluon fields
with a quark field at one end and an antiquark
field at the other, {\textit{eg}}\
$\overline{\psi}(x_1)U_{\mu}(x_1)U_{\nu}(x_1+1_{\mu})\ldots
U_{\epsilon}(x_2-1_{\epsilon})\psi(x_2)$.  Under a gauge transformation
the $G$ matrix at the beginning of one link `eats' the $G^{\dag}$ at
the end of the previous link, since $G^{\dag}G=1$. The $G$ matrices at
$x_1$ and $x_2$ are `eaten' by those transforming the quark and
anti-quark fields, if we sum over quark and antiquark colours. The same
thing happens for any closed loop of $U$s, provided that we take a
trace over colour indices. Then the $G$ at the beginning of the loop
and the $G^{\dag}$ at the end of the loop, the same point for a closed
loop, can `eat' each other. 

The purely gluonic piece of the continuum QCD action is
\begin{equation}
S_{\rm cont} = \int d^4x \frac{1}{2}  \Tr F_{\mu\nu}F^{\mu\nu}
\label{sgcont}
\end{equation}
where $F_{\mu\nu}$ is the field strength tensor,
\begin{equation}
F_{\mu\nu} = \partial_{\mu}A_{\nu}-\partial_{\nu}A_{\mu} + ig[A_{\mu},A_{\nu}].
\end{equation}
The simplest lattice discretisation of this is the so-called 
Wilson plaquette action:
\begin{equation}
S_{\rm latt} = \beta \sum_p 
\left(1 - \frac{1}{3} \Re \{\Tr U_p\}\right);\quad
\beta = \frac{6}{g^2}.
\label{sglatt}
\end{equation}
In fact the 1 here is irrelevant in the lattice QCD calculation, 
giving only an overall normalisation that 
vanishes from the ratio of path integrals, and so it is often dropped 
from $S_{latt}$.
 
\begin{figure}[ht]
\centering{
\includegraphics[width=20mm, clip, trim=0 0 0 0]{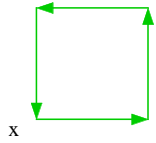}
\caption{A plaquette on the lattice.}
\label{davies:fig05}}
\end{figure}

$U_p$ is the closed $1\times 1$ loop called the plaquette, an SU(3) 
matrix formed by multiplying  4 gluon links together in a sequence. 
For the plaquette with corner $x$ in the $i,j$ plane we have (\Figure{\ref{davies:fig05}}):
\begin{equation}
U_{p,ij}(x) = U_i(x)U_j(x+1_i)U^{\dag}_i(x+1_j)U^{\dag}_j(x)
\label{Uplaq}
\end{equation}
Tr in $S_{\rm latt}$ denotes taking the trace of $U_p$ {\textit{ie}} the sum of the
3 diagonal elements. 
$S_{\rm latt}$ sums over all plaquettes of all orientations on the
lattice.  $\beta$ is a more convenient version for the lattice of the
QCD bare coupling constant, $g^2$. This is the single input parameter
for a QCD calculation (whether on the lattice or not) involving only
gluon fields. Notice that the lattice spacing is not explicit
anywhere, and we do not know its value until {\it after} the
calculation. (This is a difference from the quantum mechanical 
example of the previous section where we had to choose and input a value 
for $a$.) The value of the lattice spacing depends on the bare
coupling constant. Typical values of $\beta$ for current lattice
calculations using the Wilson plaquette action are $\beta \approx
6$. This corresponds to $a \approx$ 0.1fm. Smaller values of $\beta$
give coarser lattices, larger ones, finer lattices. 
This is obvious from the asymptotic freedom of QCD, which tells 
us that the coupling constant $g^2$ goes to zero at small distances
or, equivalently, high energies. $\beta$ is the inverse of the 
bare coupling constant at the scale of the lattice spacing and 
therefore tends to $\infty$ as the lattice spacing goes to zero. 

That $S_{\rm latt}$ of
Equation~\ref{sglatt} is a discretisation of $S_{\rm cont}$ is not
obvious, and we will not demonstrate it here. It should be clear,
however, from Equations~\ref{ulattcont} and~\ref{Uplaq} that
$S_{\rm latt}$ does contain terms of the form $\partial_{\mu}A_{\nu}$ 
needed for the field strength tensor, $F_{\mu\nu}$.

$S_{\rm latt}$ is gauge-invariant, as will be clear from earlier.
Thus lattice QCD calculations do {\it not} require gauge
fixing or any discussion of different gauges or ghost terms, 
as would be required for continuum calculations using perturbation 
theory.  
Our lattice
calculation is fully non-perturbative since the Feynman Path Integral
includes any number of QCD interactions.
In contrast to the real world, however, the calculations
are done with a non-zero value of the lattice spacing and a
non-infinite volume. In principle we must take $a \rightarrow 0$ and
$V \rightarrow \infty$ by extrapolation. In practice it suffices to
demonstrate, with calculations at several values of $a$ and $V$, that
the $a$ and $V$ dependence of our results is small, and understood,
and include a systematic error for this in our result.

Before discussing quarks we illustrate here how a lattice 
calculation is done with only gluon fields,
how the lattice spacing is determined, and 
what the systematic errors are. 

A quantity that can be calculated in the pure gluon theory 
is the expectation value of a closed loop of gluon fields. In fact 
this can be related to the QCD potential between an infinitely massive 
quark and antiquark. Although not 
directly a physical quantity, this is something about which we have some 
physical understanding. An infinitely massive quark does not move in 
spatial directions and so simply generates a path which is a 
string of gluon fields in the time direction. If this is joined to a 
string generated by an antiquark a distance $R$ in lattice units away 
then a rectangular $R\times T$ loop of gluon fields is created. This 
is known as a Wilson loop, see \Figure{\ref{davies:fig04}} right. 

To `measure' this in pure gluon QCD, we generate configurations of gluon
fields with probability $e^{-S_{latt}}$ where $S_{latt}$ is a discretisation 
of the pure gluon QCD action given by Equation~\ref{sglatt}. On each of 
these configurations we calculate the $R\times T$ Wilson loop, averaging
over all positions of it on the lattice. We then average over the 
results on each configuration in the ensemble to obtain a final answer with statistical error, 
for lots of values of $R$ and $T$. 
We have 
\begin{equation}
\frac{1}{Z}<0|{\mathcal{O}}|0> = \frac{\int {\mathcal{D}}U 
{\mathcal{O}}[U]e^{-S_{g,QCD}}}{\int {\mathcal{D}}Ue^{-S_{g,QCD}}} 
= <<{\mathcal{O}}>> = \frac{1}{N_{conf}} \sum_{i=1}^{N_{conf}} O_i
\end{equation}
Typically we need many hundreds of configurations in an ensemble for a 
small statistical error at large $R$ and $T$.

The ensemble average of $O$ is related to the heavy quark potential 
by similar arguments to those used for the operator $x(t_1)x(t_1)$ in 
section 2.1. One end of the Wilson loop creates a set of eigenstates of 
the Hamiltonian that are based on a massive quark-antiquark pair. These
eigenstates are produced with different amplitudes by the Wilson loop 
operator and have different energies. In this case there is no 
kinetic energy, so the energies are those of the heavy quark potential. 
The different eigenstates propagate for time $T$ and, if 
$T$ is large, the ground state eventually dominates. 
\begin{equation}
<<{\mathcal{O}}>> = Ce^{-aV(R)T} + C^{\prime}e^{-aV^{\prime}(R)T} + \ldots.
\end{equation}
By fitting the results as a function of the time length, $T$, in lattice 
units, the heavy quark potential in lattice units, $aV(R)$, is obtained. 
$V^{\prime}$ is some kind of excitation of the potential which we will 
not be interested in here. The heavy quark potential at short distances
should behave perturbatively and take a Coulomb form. At large distances 
we expect a `string' to develop which confines the quark and antiquark 
and gives a potential which rises linearly with separation. 
We can therefore fit the lattice potential to the form 
\begin{equation}
aV(r=Ra) = -\frac{4}{3}\frac{\alpha_s(r)}{R} + \sigma a^2 R + \tilde{C}
\end{equation}
where $\tilde{C}$ is a `self-energy' constant that appears in the lattice 
calculation. If the results for $aV(R)$ are plotted against $R$, 
the slope at large $R$ is the `string tension', $\sigma$, in lattice 
units, {\textit{ie}} $\sigma a^2$. Phenomenological models of the heavy 
quark potential give values for $\sqrt{\sigma}$ of around 
440 MeV. Using this value for $\sigma$ and the result from 
the lattice of $\sigma a^2$, gives a value for $a$. This 
is often quoted as a value for $a^{-1}$ in GeV. Note that 
$a^{-1}$ in GeV $= 0.197/(a \, {\rm in \, fm})$.

\begin{figure}[ht]
\centering{
\includegraphics[width=80mm, clip, trim=0 0 0 0]{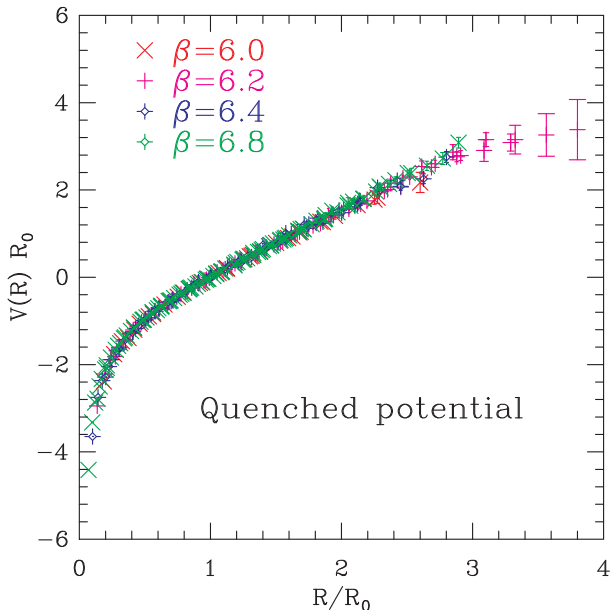}
\caption{The heavy quark potential in units of the parameter, $r_0$, 
as a function of distance, $r$, also in units of $r_0$. The 
calculations were done in quenched (pure glue) QCD at a variety 
of different values of the lattice spacing, corresponding to the 
different values of $\beta$ quoted. (Bali 2000)}
\label{davies:fig06}}
\end{figure}

The results for $aV$ can then be multiplied by $a^{-1}$ to 
convert them to physical units of GeV and, having removed the 
constant $\tilde{C}$, the results can be plotted as a function of 
the physical distance, $r$ in fm. If discretisation errors are 
small, then results at different values of the lattice 
spacing should be the same. \Figure{\ref{davies:fig06}} shows results 
for the heavy quark potential on relatively fine lattices 
at different values of $\beta$ using the Wilson plaquette 
action for $S_{g}$ (Bali 2000). $V(r)$ is not in fact given in GeV here, 
nor is $r$ in fm, but both are given  
in terms of a parameter called $r_0$ (Sommer 1994). This is the value of $r$ at 
which $r^2 \partial V/\partial r$ = 1.65 and is a commonly used 
quantity to determine the lattice spacing (or at least relative lattice 
spacings), rather than the 
string tension. $r_0$ is not a physical parameter, and as such is 
not available in the Particle Data Tables. We shall see later that 
there are good hadron masses to use for the determination of 
$a$ and these can also be used to determine the value of $r_0$. Meanwhile, 
$r_0 \approx 0.5$fm. The results at different 
values of $\beta$ lie on top of each other and this gives 
us confidence that the discretisation errors are small.  

\begin{figure}[ht]
\centering{
\includegraphics[width=110mm, clip, trim=0 0 0 0]{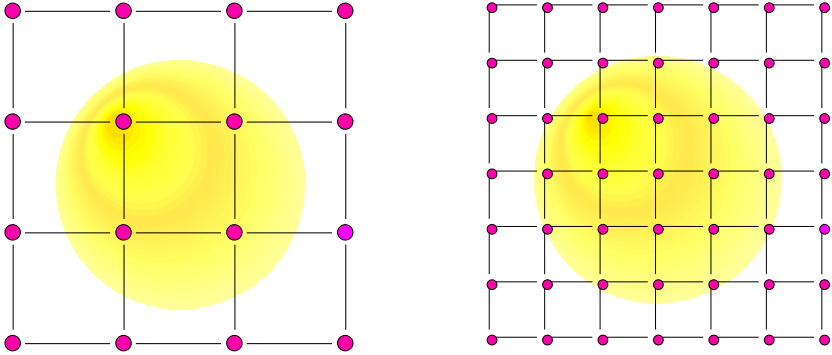}
\caption{A physical object on two lattices of different lattice spacing. 
On the right our updating algorithm takes much longer to register 
a significant change on the length scale of the object. }
\label{davies:fig07}}
\end{figure}

The values of $\beta$ used correspond to rather fine lattices. 
The coarsest lattice is at $\beta$ = 6.0 and has $a$=0.1fm. 
The finest lattice is at $\beta$ = 6.8 and has $a$ = 0.03fm. 
The finest configurations 
are very expensive to generate. Yet if we looked closely at the results 
at $\beta$ = 6.0 we would be able to see discretisation 
errors at the few percent level. If 
we want accurate results on lattices that are coarse enough for 
affordable calculation (especially, as we shall see, once 
we include quarks) then we must improve the discretisation of 
the action. The cost of lattice calculations grows naively as 
$a^{-4}$ because of the four-dimensional space-time. In fact 
it is even worse than this because of critical slowing-down. 
Physical distances on the lattice grow in lattice units as 
the lattice spacing gets smaller (see \Figure{\ref{davies:fig07}}). This means that algorithms 
that update configurations by making local changes to the fields, 
get slower and slower at making a change on a distance scale 
of $r$ as $a$ is reduced. Then the cost of 
the calculation of, say, $V(r)$ actually grows as $a^{-6}$.  
It becomes imperative to improve the action, rather than to 
attempt to beat down the discretisation errors by reducing 
$a$. 

Improving the gluon action is at first sight straightforward. 
We expand the Wilson plaquette action in powers of $a$ and 
notice that it has $a^2$ errors when compared to the continuum 
QCD gluon action. 
We add a higher dimension operator to the action in the 
form of a $2\times 1$ Wilson loop to cancel this error: 
\begin{equation}
S_{g,latt} = \beta \sum (1 - \frac{5c_1}{3}\frac{{\rm Re Tr} U_p}{3} + \frac{c_2}{12}\frac{{\rm Re Tr} 
(U_{2\times 1}+U_{1\times 2})}{3}).
\end{equation}
$c_1$ and $c_2$ are chosen to remove the $a^2$ error 
and naively would have the value 1. However, in a quantum field theory like 
QCD, the value of $c_{1,2}$ becomes renormalised by radiative corrections. 
$S_{g,latt}$ must reproduce $S_{g,cont}$ to a required level of accuracy and so 
$c_{1,2}$ must absorb the effect of the differences in gluon radiation 
between the continuum and lattice versions of QCD. This difference arises 
from gluon radiation with momentum larger than $\pi/a$ which does 
not exist on the lattice. Gluon radiation at these high momenta is 
perturbative and so we can calculate $c_{1,2}$ as a perturbative 
power series in $\alpha_s$. Provided $a$ is small enough so that 
$\alpha_s(\pi/a)$ is small enough, the $c_{1,2}$ will not be very different from 1. 
There is a complication, however, and that arises from the 
way in which the lattice gluon field $U_{\mu}$ is related to the usual  
gluon field $A_{\mu}$. The exponential relationship in Equation~\ref{ulattcont}
means that the lattice perturbation theory contains vertices with many 
powers of $A_{\mu}$. Although these are suppressed by powers of $g$, they produce rather 
large contributions and have to be taken into account. In fact they 
appear as so-called `tadpole' diagrams which take the same form in 
many different processes (Lepage and Mackenzie 1993). 
This allows them to be substantially removed in a universal way by the simple 
expedient of estimating how far the gluon link field, $U_{\mu}$, is from the 
value 1 (the $3\times 3$ unit matrix) that it  would take in the 
continuum, $g^2 \rightarrow 0$ limit. 
We can measure this, for example, from the average 
plaquette (traced and divided by 3 so that it would be 1 in the 
continuum limit). This contains 4 $U$ fields so we take the fourth 
root to determine the `tadpole-factor', $u_0$. 
Taking
\begin{equation}
U_{\mu}(x) \rightarrow \frac{U_{\mu}(x)}{u_0}
\end{equation}
is called `tadpole-improvement'. If this is done then the calculation 
of $c_{1,2}$ does indeed give well-behaved perturbative expansions 
which do not differ substantially from the naive value of 1. The 
improved gluon action becomes
\begin{equation}
S_{g,latt} = \beta \sum ( - \frac{5c_1}{3}\frac{{\rm Re Tr} U_p}{3u_0^4} + 
\frac{c_2}{12}\frac{{\rm Re Tr} (U_{2\times 1}+U_{1\times 2})}{3u_0^6}),
\label{tadimpglue}
\end{equation}
dropping the constant term in the action. Taking $c_{1,2}$ to be 1 is the 
tree-level tadpole-improved Symanzik action and this gives results significantly 
better than the Wilson plaquette action on coarse lattices (Alford \etal 1995, Lepage 1996).
Note that $\beta$ here cannot be compared directly to that for the Wilson plaquette 
action. It only makes sense to compare results between different actions at 
the same value of the lattice spacing. 

\begin{figure}[h!]
\centering{
\includegraphics[height=45mm, clip, trim=6 0 0 10]{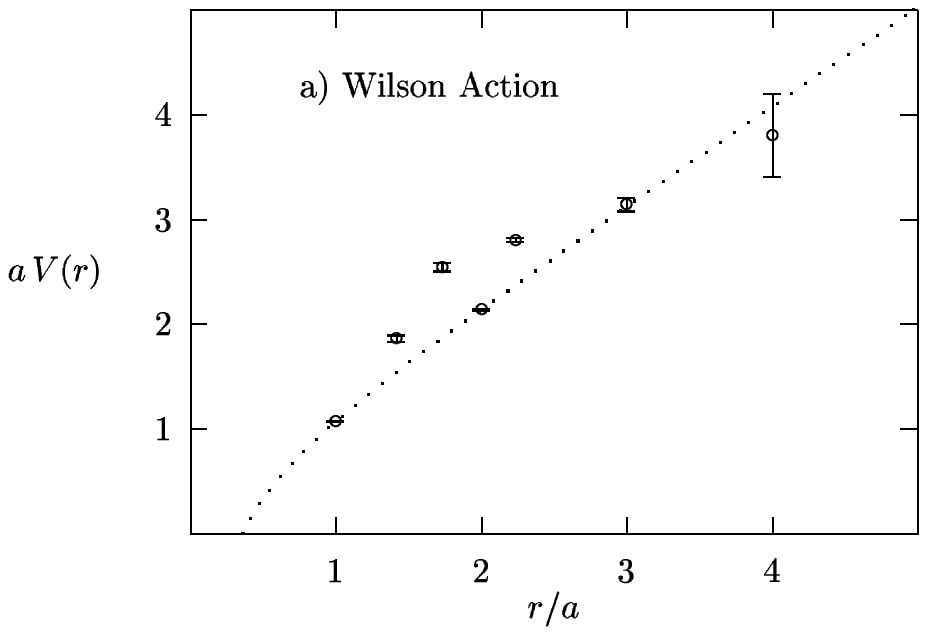}
\includegraphics[height=46mm, clip, trim=40 20 10 0]{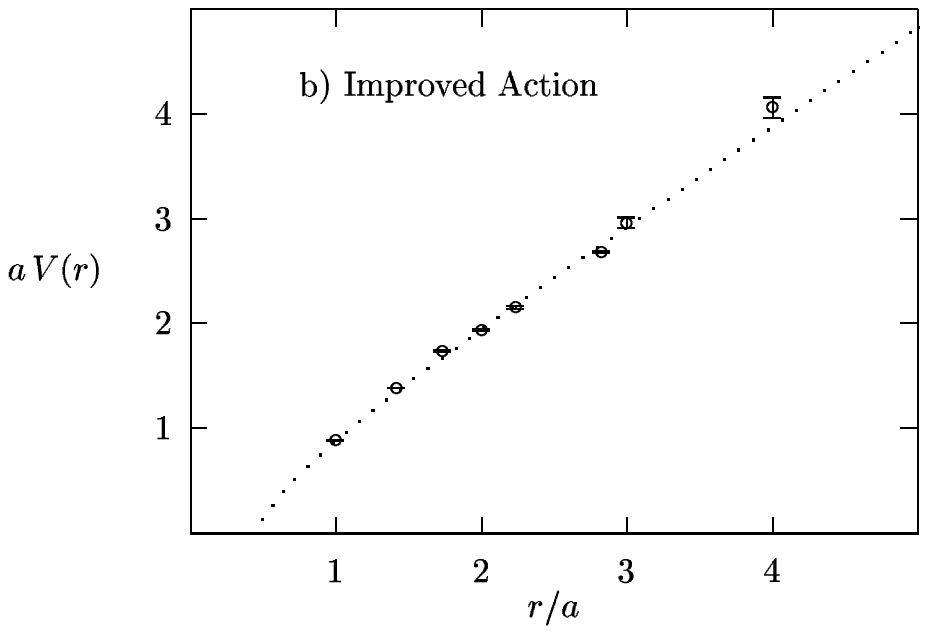}
\caption{The heavy quark potential calculated in pure gluon 
QCD at a lattice spacing, $a$, of 0.4fm. The left plot uses the 
Wilson plaquette action, the right plot the tree-level 
tadpole-improved Symanzik action. The right plot has a much 
smoother curve than the upper one reflecting the 
improvement of discretisation errors (Alford \etal 1995).}
\label{davies:fig08}}
\end{figure}

\Figure{\ref{davies:fig08}} compares results for the heavy quark potential,
calculated as discussed above, for the two actions on coarse lattices with 
lattice spacing 0.4fm (Alford \etal 1995). The results with the Wilson plaquette action 
show clear discretisation errors in the fact that $V(r)$ is not a smooth 
curve, {\textit{ie}} it does not have rotational invariance. 
It reflects the fact that distances on the lattice are not in general 
travelled in straight lines and the result will depend on the 
actual path used between two points. 
The curve obtained with the improved gluon 
action is much smoother. As discretisation errors are removed, 
the path dependence of the distance measurement is reduced and
the rotational invariance of the continuum is restored. Indeed $c_{1,2}$ could be fixed 
non-perturbatively ({\textit{ie}} within the lattice simulation) 
by demanding rotational invariance and it is clear 
from this that results close to 1 after tadpole-improvement would 
be obtained. 

In section 3 we will describe recent lattice results and these 
use an improved gluon action to reduce the discretisation errors. 
In fact the action is improved beyond that used in \Figure{\ref{davies:fig08}}
by including the first $\alpha_s$ terms in $c_1$ and $c_2$ and an additional 
operator that appears at $\cal{O}$$(\alpha_s)$ made of a 3-dimensional 
parallelogram of $U$ fields (Alford \etal 1995, Lepage 1996).  

\subsection{Quark fields in Lattice QCD}

Quarks represent a big headache for Lattice QCD. Because quark fields anticommute, 
they cannot be handled using ordinary numbers on a computer.
We therefore have to perform the integral over the quark fields in the 
path integral by hand. In fact this is easy to do because of the way that 
the quark fields appear in the QCD action. Including the Dirac piece 
of the action for quarks, we have 
\begin{eqnarray}
Z &=& \int {\cal{D}}U\,{\cal{D}}\psi\,{\cal{D}}\overline{\psi}\,e^{-S_{QCD}} \nonumber \\\
S_{QCD}
&=& S_{g} + \sum_x\overline{\psi}(\gamma\cdot D+m)\psi = S_{g} + \overline{\psi}M\psi.
\end{eqnarray}
The quark field, $\psi$ is a 12-component (3 for colour and 4 for spin) field 
at every lattice point, so $M$ is a matrix with $12\times L^3T$ rows and columns 
for an $L^3T$ lattice. $D$ is a covariant derivative that includes a coupling 
with the gluon field.  
The integral over quark fields gives
\begin{equation}
Z = \int {\cal{D}}U\,{\rm det}(M)\,e^{-S_{g,QCD}} = \int {\cal{D}}U\, e^{-\tilde{S}_{QCD}} 
\end{equation}
where $\tilde{S} = S_{g,QCD} +\ln{\rm det}(M)$, one det$M$ factor per quark flavour. 

\begin{figure}[ht]
\centering{
\includegraphics[width=60mm, clip, trim=0 0 0 10]{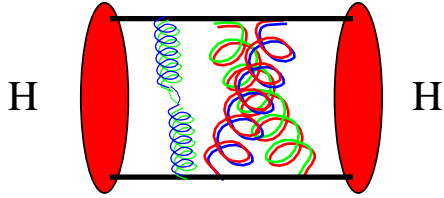}
\caption{An illustration of the calculation of a simple 
2-point function for a hadron in lattice QCD.}
\label{davies:fig09}}
\end{figure}

Typically in lattice QCD we want to calculate the mass of a hadron
made of quarks and antiquarks. To do this we must 
calculate the expectation value of a product of appropriate operators
made of quarks and antiquarks. A suitable operator that creates
a meson is $\overline{\psi}^{a,\alpha,f_1}\Gamma^{\alpha\beta}{\psi}^{a,\beta,f_2}$ 
where $\Gamma$ is some combination of $4\times 4$ $\gamma$ matrices
that give the right spin-parity ($J^P$) quantum numbers to the meson,
$a$ is a colour index, $\alpha$ and $\beta$ are spin indices and $f_i$ 
are flavour indices. We include the conjugate operator to 
destroy the meson $T$ lattice spacings away in time. Then, suppressing colour and spin, 
\begin{equation}
\frac{1}{Z}\langle 0 | H^{\dag}(T)H(0) | 0 \rangle = 
\frac{\int {\cal{D}}U\,{\cal{D}}\psi\,{\cal{D}}\overline{\psi}\,
(\sum_{\vec{x}}\overline{\psi}^{f_1}\Gamma\psi^{f_2}(\vec{x}))_T(\overline{\psi}^{f_2}\Gamma\psi^{f_1})_0 
e^{-S_{QCD}}}{\int {\cal{D}}U\,{\cal{D}}\psi\,{\cal{D}}\overline{\psi}\,e^{-S_{QCD}}}.
\end{equation}
On integration, the right-hand-side becomes (for the case where $f_1 \ne f_2$):
\begin{equation}
\frac{\int {\cal{D}}U\,{\rm Tr_{spin,colour,\vec{x}}}(M^{-1}_{f_1(0,T)}\Gamma M^{-1}_{f_2(T,0)}\Gamma )
{\rm det}M\,e^{-S_{g,QCD}}}{\int {\cal{D}}U\,{\rm det}M \, e^{-S_{g,QCD}}}.
\label{hadprop}
\end{equation}
This calculation requires us to generate sets of gluon field configurations 
with probability $e^{-\tilde{S}_{QCD}}$, calculate the 
trace over spin and colour of the $M^{-1}$ factors on each 
configuration and then average these over the ensemble. The calculation  
is illustrated in \Figure{\ref{davies:fig09}}. It is known as a 2-point 
function calculations because there are two operators at 
times 0 and $T$, represented by the filled ovals. The straight lines 
at the top and bottom indicate the valence quark propagators (the 
$M^{-1}$ factors above that connect the creation and annihilation operators 
for that particular valence quark). As the quark propagates it 
interacts any number of times with the other quark through the lattice 
gluon field and this is indicated by the curly lines. Some of these 
gluon lines may include the effect of a sea quark-antiquark pair (as 
on the left of the diagram). 

If we have $f_1 = f_2$ for a flavour-singlet quantity then there are 
additional `disconnected' pieces where the $M^{-1}$ factors are generated 
from each operator separately. This gives, for example, 
${\rm Tr}(M^{-1}_{(0,0)}){\rm Tr}(M^{-1}_{(T,T)})$. These are very difficult 
to calculate, being very noisy, and accurate results for light flavour-singlet 
mesons are not yet available. For heavy mesons, $c\overline{c}$ and $b\overline{b}$, 
this is expected to be a very small effect and is usually ignored.  

The hadron mass is determined from the results through the 
usual multi-exponential form 
\begin{equation}
\frac{1}{Z}\langle | 0 H^{\dag}(T)H(0) | 0 \rangle = << 
{\rm Tr_{spin,colour,\vec{x}}}(M^{-1}_{f_1}\Gamma M^{-1}_{f_2}\Gamma ) >> = \sum_n C_n e^{-m_naT}.
\label{multifit}
\end{equation}
$n$ runs over radial excitations of the hadron with a particular 
set of $J^P$ and flavour quantum numbers, and $m_na$ are the different 
hadron masses. In fact the formula above is only correct for a lattice 
of infinite time extent in lattice units, $T^{\prime}$. On the finite 
lattices we must use there are additional terms from the possibility 
of quarks `going backwards' round the lattices. These terms take 
the form $e^{-m_na(T^{\prime}-T)}$ and can readily be taken account 
of in the fit, although their form does depend on the hadron being 
studied. $C_n$ in the above equation is related to the square of the matrix 
element $<0|H|n>$ by analogy with Equation~\ref{x1x2} in section 2.1.   
The size of $C_n$ then depends on the form of the operator 
$H$ used to create and destroy the hadron. Any operator with the 
right quantum numbers can be used and if we make a set of 
such operators we can calculate a whole matrix of correlators 
and fit them simultaneously for improved precision. Typically we use 
operators made of quark and antiquark fields but separated in space
according to some kind of `wavefunction', (known as 
a `smearing') {\textit{eg}} $\overline{\psi}_{x}\phi(x-y)\psi_y$. To make 
a gauge-invariant operator of this kind either requires strings of 
$U$s to be inserted between $x$ and $y$ or for the gluon field 
configuration to be gauge-fixed, typically to Coulomb gauge. 
It may be possible to choose $\phi$ so that a particularly good overlap 
with one of the states of the system ({\textit{eg}} the ground state) is obtained, 
and poor overlap with the others. Then $C_0$ would be large in 
the equation above, and the other $C_n$ small, and this would 
give improved precision for $m_0a$. The masses of the radial 
excitations are of interest too and smearings which improve 
overlap with them have also been studied. With particular 
operators, $H$, $C_n$ contains information of physical 
relevance. For example the matrix element of the temporal 
axial current, $J_{A_0}$ between the vacuum and a pseudoscalar 
meson gives its decay constant, related to the rate for its 
decay purely to leptons via a $W$ boson. The calculation in 
lattice QCD of matrix elements of this kind is discussed 
in section 3.3.  

The way in which a lattice QCD calculation with quarks is done 
makes a clear distinction between valence quarks and 
sea quarks. The valence quarks are those that give the hadron 
its quantum numbers and these give rise to the $M^{-1}$ factors 
in Equation~\ref{hadprop}. The sea quarks are those that 
are produced as quark-antiquark pairs from energy fluctuations 
in the vacuum (quark vacuum polarisation). They give rise to the det$M$ factors in 
Equation~\ref{hadprop}. The important sea quarks are the light ones, 
$u$, $d$ and $s$. The heavy quarks $c$ and $b$ have no effect as 
sea quarks and the $t$ quark does not even have to be considered 
as a valence quark since it does not form bound states before 
decaying. 

Manipulations of the matrix $M$ are computationally costly. Even though 
it is a sparse matrix (with only a few non-zero entries) it is very 
large. There are various computational techniques for calculating 
the $M^{-1}$ factors and the det$M$ factor is included by repeated 
determination of the calculation of $M^{-1}$. This makes the 
inclusion of det$M$ very costly indeed. It becomes increasingly 
hard as the quark mass becomes smaller because $M$ becomes 
ill-conditioned. (The eigenvalues 
of $M$ range between some fixed upper limit and the quark mass, 
so this range increases as $m_q \rightarrow 0$).
In the real world the $u$ and $d$ quarks have very small mass 
and so in the past there has not been sufficient computer power 
available to include them as sea quarks or, if they have been included, their 
masses have been much heavier than their real values. 

Missing out sea quarks entirely is known as the quenched 
approximation. It is clearly wrong, but for a long time the 
presence of other systematic errors and poor statistics obscured 
this fact. More recently it has become clear that the systematic 
error in the quenched approximation is around 10-20\%. 
When light quark vacuum polarisation (det$M$) is included 
the calculation is said to be `unquenched' or `dynamical'. 
The sea quarks are then also called dynamical quarks. 
We will see in the results section that it is now possible 
to include realistic quark vacuum polarisation effects and 
the quenched approximation can be laid aside at last. 

Our earlier discussion makes clear that the computational 
cost of including light quark vacuum polarisation will 
require a very good ({\textit{ie}} highly improved) discretisation 
of both the gluon and quark actions. We have discussed the 
gluon action earlier. For the quark action there are  
several possibilities, or formulations. We will concentrate 
here on those formulations that have already been used in 
unquenched simulations. 

The naive quark action is a straightforward discretisation 
of the Dirac action in Euclidean space-time:  
\begin{equation}
S_q = \sum_x \overline{\psi}(x)(\gamma\cdot\Delta + ma) \psi (x).
\end{equation}
The finite difference $\Delta$ is
\begin{equation}
\Delta_{\mu}\psi(x) \equiv 
\frac{1}{2}(U_{\mu}(x)\psi(x+1_{\mu}) - U^{\dag}_{\mu}(x-1_{\mu})\psi(x-1_{\mu}))
\end{equation}
showing explicitly how the gluon fields appear, coupled to the quarks, to 
maintain gauge invariance in 
the action. QCD with quarks has parameters, in addition to the coupling 
constant, that are the quark 
masses. In lattice QCD these appear as quark masses in lattice units, $ma$.
As with the lattice spacing/coupling constant, the quark masses are not known a priori and 
must be adjusted as a result of a lattice calculation. We therefore iterate until, 
for example, we get a particular hadron mass correct. Since the quark mass is 
in lattice units, finer lattices will require smaller values of $ma$ than 
coarser ones.   

The naive quark action has several good properties. It has 
discretisation errors at $\cal{O}$$(a^2)$.  
The finite difference piece is anti-hermitian and therefore has purely 
imaginary eigenvalues that appear as $\pm i\lambda$. With the 
mass term the eigenvalues of $M$ are then $ma \pm i\lambda$. This 
is the same as in the continuum and follows from the chiral symmetry 
of the action {\textit{ie}} our ability to rotate independently left- and 
right- handed projections of the quark field. It means that the bare 
quark mass is simply multiplicatively renormalised and the 
renormalised quark mass goes to zero at the same point as the bare 
quark mass. This is important because we need to work with very 
small (renormalised) quark masses for the $u$ and $d$ quarks and we want to be able 
to go in that direction by just taking $ma$ to be smaller and 
smaller (if we have enough computing power).  

Chiral symmetry is 
spontaneously broken in the real world giving rise to a Goldstone boson, the $\pi$, 
whose mass consequently vanishes at zero quark mass ($m_{\pi}^2 \propto m_q$). For small 
quark masses, where $m_{\pi}$ is small, we have a well-developed 
chiral perturbation theory which tells us how hadron masses 
and properties should depend on the $u/d$ quark masses (or 
equivalently $m_{\pi}^2$) and we can make use of this to extrapolate 
down to physical $u/d$ quark masses from the results of our lattice 
QCD simulation provided that we are able to work at small enough 
$u/d$ quark masses to be in the regime where chiral perturbation 
theory works. In general $m_{u/d} < m_s/2$ is necessary for an accurate 
extrapolation. We will discuss this further in section 3. (Arndt 2004)   

\begin{figure}[ht]
\centering{
\includegraphics[height=45mm, clip, trim=0 0 0 0]{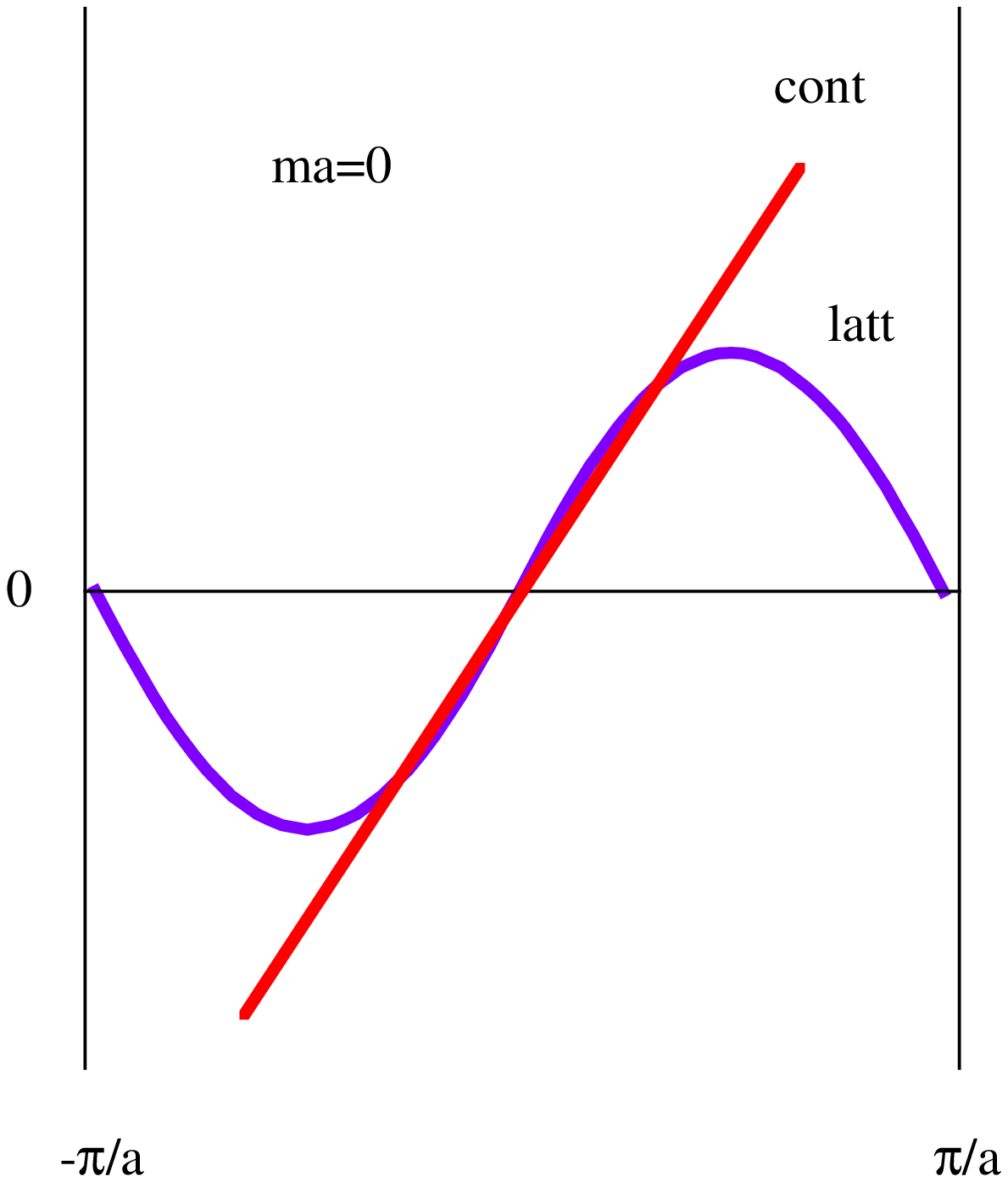}
\includegraphics[height=40mm, clip, trim=-60 0 0 0]{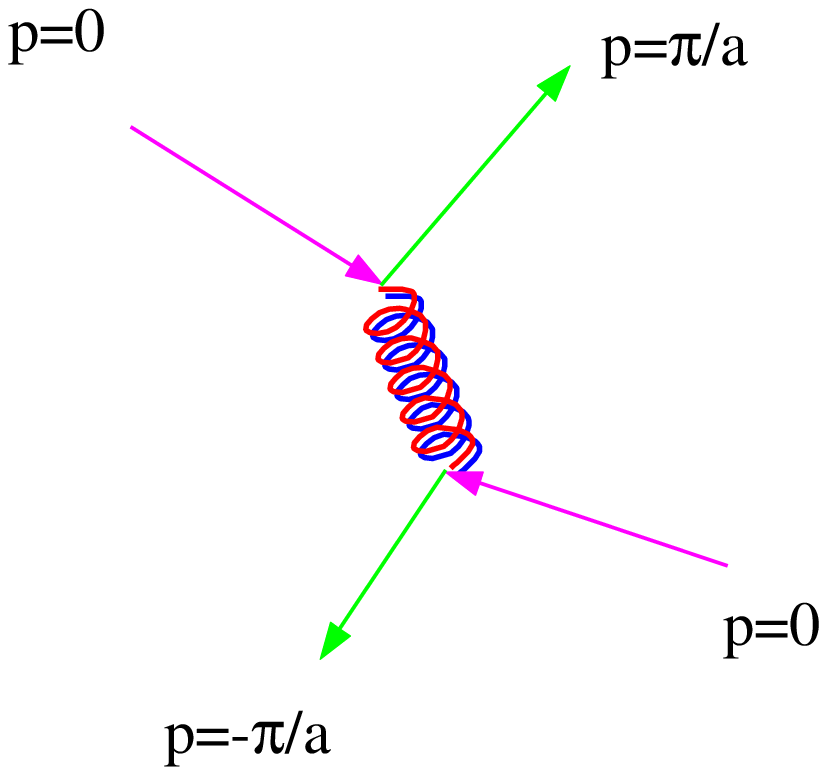}
\caption{(Left) The continuum derivative and lattice finite 
difference compared as a function of momentum within the 
first Brillouin zone on the lattice. The points 
$p = \pi/a$ and $-\pi/a$ are joined by lattice periodicity. 
(Right) Taste-changing interactions for naive quarks. A high-momentum 
($p=\pi/a$) gluon exchange can change one taste of quark into 
another.}
\label{davies:fig10}}
\end{figure}

It would seem that we have everything ready to do the simulations but 
we must first discuss the infamous doubling problem. The naive 
quark action in fact describes 16 quarks rather than 1. This is 
demonstrated most easily in 1-dimension in the absence of gluon fields.
It suffices to compare the continuum derivative in momentum space, 
{\textit{ie}} $p$, with the Fourier transform of the finite difference on 
the lattice, which is $\sin(pa)/a$. These are plotted in \Figure{\ref{davies:fig10}}
as a function of $p$ between $-\pi/a$ and $\pi/a$, the limits 
over which $p$ makes sense on the lattice (the first Brillioun zone). 
Momenta larger than $\pi/a$ reappear as negative momenta larger 
than $-\pi/a$ so these points are periodically connected. 
Around $p \approx 0$ the sine function goes through zero and mimics 
a straight line up to $a^2$ errors 
as expected. The problem is that this is also true around $p \approx \pi/a$
(with opposite slope). This means that there is a continuuum-like solution 
of the Dirac equation around $p = \pi/a$ as well as around $p = 0$
{\textit{ie}} there are 2 quarks instead of 1. In 4-d (and including $\gamma$ 
matrix algebra as well) the picture is more complicated but the 
basic facts are the same - there are $2^4$ quarks instead of 1. 
The `doublers' or extra `tastes' of quark live at the edges of 
the Brillouin zone, where any component of $p$ is close to $\pi/a$. 

This was originally thought to be disastrous and huge efforts, which 
continue today, were made to solve the problem. Wilson introduced 
a Wilson term into the naive action which has the effect of 
giving mass to the doublers. The Wilson action is then 
\begin{eqnarray}
S_f^W &=& S_f^{naive} - \frac{r}{2} \sum_x \bar{\psi}_x \Box \psi_x  \nonumber \\
\Box \psi_x &=& \sum_{\mu=1}^4 U_{\mu}(x){\psi}_{x+1_{\mu}} - 2 \psi_x + 
U^{\dag}_{\mu}(x-1_{\mu})\psi_{x-1_{\mu}}. 
\end{eqnarray}
In the absence of gluon fields the Fourier transform of the extra term is 
$2r\sin^2(pa/2)$ which has a maximum at $\pm \pi/a$. When folded in to the 
inverse quark propagator, taking account of the $\gamma$ matrix structure, 
this then prevents the inverse propagator from vanishing at the edges of the 
Brillouin zone, effectively giving the doublers a large mass which
increases as $a \rightarrow 0$ so that they are completely removed 
in the continuum limit. 

This solution of the doubling problem was used for many years. It does have 
two big disadvantages. One is that discretisation errors now appear 
at $\cal{O}$$(a)$ which causes large systematic errors even on relatively 
fine lattices. The other is that the Wilson term does not respect 
chiral symmetry and so we lose the simple connection between the bare quark mass parameter
and the renormalised quark mass and have to search in $ma$ for 
the point where $m_{\pi}$ vanishes. The eigenvalue spectrum is now 
complicated as well which creates numerical difficulties. 
Discretisation errors are ameliorated in the improved form of the action 
known as the `clover' action. The $\cal{O}$$(a)$ errors are cancelled 
by a $\sigma_{\mu\nu}F^{\mu\nu}$ term which is discretised as a set 
of plaquettes around a central point that looks like a 4-leaf clover. 
The coefficient of the clover term can be set using tadpole-improvement or 
non-perturbatively. Again it is close to 1 {\it after} the tadpole-improvement
with a well-behaved perturbative expansion. 
Work using the clover action continues and it is being used for 
unquenched simulations (Ishikawa \etal 2005). A newer form of the action, called twisted mass 
QCD, is also being developed. This overcomes several of the numerical 
problems and is a promising development for the future (Frezzotti 2005). 

Another relatively new development is that of a set of actions that 
maintain chiral symmetry on the lattice but also solve the doubling 
problem. These require an enormous increase in computer power because 
they either need a matrix inversion inside a matrix inversion to 
calculate $M^{-1}$ (overlap quarks) or a calculation in 5-dimensions 
(domain wall quarks). Calculations using these formalisms have then been 
restricted to the quenched approximation because of the cost. In future 
they may be possible to implement including realistic quark vacuum polarisation 
effects on very powerful supercomputers (Chiu 2004). 

The recent progress in lattice QCD calculations has come, however, 
from returning to the naive discretisation of the quark action. 
In fact the doubling problem is not a problem if the doublers are 
simply copies of each other, because we can then simply `divide by 16'
inside our calculation. If $M$ has a 16-fold degeneracy in its 
eigenvalue spectrum then taking $({\rm det} M)^{1/16}$ gives us the 
effect of 1 taste for each flavour. The problem is that the different tastes of quarks 
do interact with each other and change taste and this splits some of 
the degeneracy. The process by which this 
happens is a high momentum ($\pi/a$) gluon exchange as in \Figure{\ref{davies:fig10}}
which moves the quark from one Brillouin zone boundary to another. 
These taste-changing interactions are an artefact of using the lattice and 
they induce large discretisation errors, even though they are formally 
$\cal{O}$$(a^2)$ (Lepage 1998b). To improve the naive action then requires 
removing the unphysical taste-changing interactions at leading order 
as well as the usual $a^2$ errors from discretising a derivative as 
a symmetric finite difference, discussed in section 2.1. 
This markedly improves the discretisation 
errors, but also significantly reduces the taste-changing interactions, 
and makes this action a good one for lattice simulations.

The naive quark formalism is fast numerically because it 
can be converted to the staggered quark formalism in which there is 
no quark spin. It is a remarkable feature that the 
quark fields can be rotated so that the 
naive quark action (or its improved variant) is diagonal in spin space. 
We have only to simulate one spin component on the lattice 
and then, if required, we can reconstruct all quantities in 
terms of naive quarks. If we take 
\begin{equation}
\psi(x) \rightarrow \Omega(x)\chi(x); \quad \bar{\psi}(x) \rightarrow \bar{\chi}(x)\Omega^{\dag}(x)
\end{equation}
where
\begin{displaymath}
\Omega(x) \equiv \prod_{\mu=0}^{3}(\gamma_{\mu})^{x_{\mu}},
\end{displaymath}
the unimproved naive action becomes
\begin{displaymath}
\bar{\psi}(x)(\gamma\cdot\Delta + ma)\psi = \bar{\chi}(x)(\alpha(x)\cdot\Delta + ma)\chi(x)
\end{displaymath}
where $\alpha(x)$ is diagonal in spin space, 
$\alpha_{\mu}(x) = (-1)^{x_0+\ldots x_{\mu-1}}$.
Rewriting the action in terms of a single-spin (but 3-colour) component field $\chi$, 
we have the unimproved staggered quark action:
\begin{equation}
S_f^S = \sum_x \bar{\chi}_x \{ \frac{1}{2} \sum_{\mu} \alpha_{x,\mu}(U_{\mu}(x)\chi_{x+1_{\mu}}
- U_{\mu}^{\dag}(x-1_{\mu})\chi_{x-1_{\mu}}) 
+ ma \chi_x \}.
\label{staggaction}
\end{equation}

The staggered quark action does not solve the doubling problem, but removes 
an exact 4-fold degeneracy from it. There are then 4 tastes for every quark flavour 
that we include with $M$ instead of 16. If we improve the action so 
that the $a^2$ taste-changing 
interactions are small then the tastes will be close to being copies of each other. 
We will expect to see a 4-fold close-to-degeneracy in the eigenvalue spectrum 
of the improved staggered $M$, which will become closer and closer as $a \rightarrow 
0$. This has been demonstrated numerically (Follana \etal 2004) 
and justifies `dividing by 4' by taking 
$({\rm det}M)^{1/4}$ in the path integral to include one quark flavour. 
The improvement of the staggered quark action to remove the leading-order 
taste-changing interactions of \Figure{\ref{davies:fig10}} is done by replacing 
the gluon fields that appear in the finite difference of Equation~\ref{staggaction}
with a combination of gluon fields that remove the coupling between 
the quark and a single $p=\pi/a$ gluon. The combination `smears' out the gluon field
in directions perpendicular to its link by including in combination with a 
single $U_{\mu}(x)$ paths such as `the staple', $U_{\nu}(x)U_{\mu}(x+1_{\nu})U^{\dag}_{\nu}(x+1_{\mu})$. 
The simplest improved staggered action with $a^2$ errors removed 
uses paths up to those containing 
7 $U$ fields in all 3 directions perpendicular to a given link
with tadpole-improvement. It is called the asqtad action and has 
been successfully used in extensive unquenched simulations 
as described in section 3 on results (Lepage 1998b, Orginos \etal 1999). 

\begin{figure}[ht]
\centering{
\rotatebox{270}{\includegraphics[width=100mm, clip, trim=0 0 0 150]{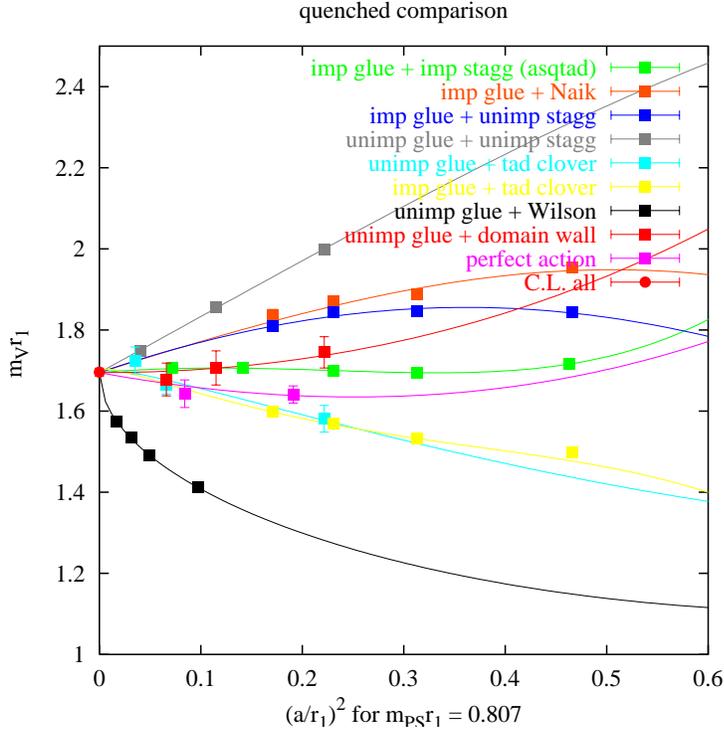}}
\caption{Results for the vector light meson mass at fixed physical 
quark mass in quenched lattice QCD using a variety of gluon 
and quark actions (for the valence quarks). The mass is given in 
units of $r_1$ and plotted 
as a function of $a^2$, also in units of $r_1$.
The lines represent fits to the lattice results, 
including appropriate discretisation errors for each case, that require 
all lines to have a common continuum limit. A good fit is obtained 
(Davies \etal 2005).}
\label{davies:fig11}}
\end{figure}

All quark formulations must give the same physical results in the 
continuum $a \rightarrow 0$ limit and this can be usefully 
checked in the quenched approximation. \Figure{\ref{davies:fig11}} shows 
results for the mass of a vector light meson (a `$\rho$') 
at an arbitrary fixed physical quark
mass as a function of lattice spacing (Davies \etal 2005). Both axes are given in 
units of $r_1$, defined in a similar to way to $r_0$, but $r_1 \approx$ 0.35fm. 
The $x$ axis is 
in terms of the squared lattice spacing since most of 
the formalisms plotted have discretisation errors at 
$\cal{O}$$(a^2)$ or better. The Wilson formalism, with 
errors at $\cal{O}$$(a)$, is clearly much worse than 
the other formalisms on this plot, with a very steep $a$ dependence, 
but it does give a consistent continuum limit.  

Heavy quarks ($b$ and $c$) represent a rather different set of issues to 
those for light quarks in lattice QCD. They have large quark 
masses, $m_Q$, and therefore large values of $m_Qa$ at any value of 
$a$ at which we are able to do QCD simulations. This means 
that we risk large discretisation errors when handling these quarks 
if we use a formalism in which the errors are set by the size of $ma$. 
If $m_Qa > 1$ then 
no amount of improvement will give a good discretisation 
for these quarks. This is particularly true for the $b$ quark 
which has a mass of around 5 GeV. To reduce $m_Qa$ below 1 requires $a^{-1} > $ 
5 GeV or
a lattice spacing $<$ 0.04 fm which is incredibly expensive 
to simulate (as well as being wasteful). Luckily physical understanding of these systems comes
to our rescue here. The experimental spectrum of hadrons containing 
$b$ and $c$ quarks shows clearly that the masses of these hadrons 
are much larger than the differences in mass between the hadrons 
in different radial and orbital excitations (and these differences 
are in fact very similar for $b$ and $c$ systems). 
The differences reflect typical kinetic energies and momenta 
inside the hadrons of a few hundred MeV to a GeV, {\textit{ie}} of the 
same size or somewhat larger than 
those typically inside light hadrons but 
$<< m_Q$. The quark mass itself is not important for the 
dynamics of the bound state
but simply provides an 
overall mass shift (Davies 1998). If we formulate the quark action 
in a way that simulates accurately nonrelativistic
quark momenta and kinetic energies we will be able to 
accurately determine the properties of the hadrons and
it is these scales that 
will also control the discretisation errors. 

This means using a nonrelativistic effective field theory and 
the example we give here is Nonrelativistic QCD, NRQCD. 
The lattice NRQCD action is a discretised nonrelativistic 
expansion of the Dirac action, matched to continuum 
QCD to the desired order in the heavy quark velocity, $v_Q$ and 
in the strong coupling constant, $\alpha_s$ (Lepage \etal 1992). 
The first few terms of the continuum NRQCD quark Lagrangian are:
\begin{equation}
{\cal{L}}_Q = \overline{\psi} ( D_t - \frac {\vec{D}^2} {2m_Q} - 
c_4 \frac {\vec{\sigma}\cdot \vec{B}} {2m_Q} + \ldots ) \psi 
\end{equation} 
where we include just the non-relativistic kinetic energy term and the 
coupling between the quark spin and chromomagnetic field. Higher order 
terms give the spin-orbit interaction and Darwin term {\textit{{\textit{etc.}}}} 
$\psi$ is a 2-component spinor here but the antiquark 
Lagrangian is simply related to the quark one and so the 
antiquark propagator is easily determined without a separate 
calculation. 
On the lattice the covariant time and space derivatives 
above become finite differences with $U$ fields included. The presence of a 
single time derivative means that the calculation of the 
quark propagator, $M^{-1}$, can be solved as 
an initial value problem on one pass through the lattice. This is much 
faster than the iterative methods needed for light quarks. 
On the lattice $m_Q$ becomes $m_Qa$, the quark mass in lattice units, and 
this must be determined by getting a heavy hadron mass correct, for 
example the $\Upsilon$ mass. This is slightly more complicated than 
in the light quark mass case because the direct mass term is missing from 
the Lagrangian, so the energy exponent of an $\Upsilon$ correlator  
at zero momentum will not give the mass (although energy differences do 
give mass differences). Instead we have to calculate hadron energies at
non-zero spatial momentum and extract the mass of the hadron from its kinetic energy.

NRQCD contains, by design, the low momentum physics of QCD for 
heavy quarks. High-momentum interactions of continuum QCD are 
missing, and the lattice in any case does not allow for $p > \pi/a$. 
The effect of these missing high-momentum interactions 
is included in NRQCD through a renormalisation of the coefficients, 
$c_i$, of the higher order terms. Because the effects that we are 
talking about are high momentum, the calculation can be done in 
perturbation theory as a powers series in $\alpha_s$. Again small 
deviations from 1 are found after tadpole-improvement of 
the $U$ fields is implemented, provided $m_Qa$ is not large. The 
$c_i$ contain functions of $m_Qa$, including inverse powers of $m_Qa$,
 multiplying powers of $\alpha_s$ and so we cannot take $a$ to 
zero or we lose control of the convergence of the $c_i$. 
This is not a problem in practice since, as discussed earlier, 
discretisation errors can now be controlled in this formalism 
to high accuracy at values of the lattice spacing that we can 
simulate and there is no need to take $a$ very small. 

For charm quarks the situation is not as clear as for $b$ quarks 
because $m_ca \approx 1$ on lattices in current use. NRQCD does not 
necessarily work well and it may be more advantageous to take 
a relativistic formalism of the $c$ quark and attempt to improve 
it highly to reduce the discretisation errors in $m_ca$ to a 
reasonable level. The Fermilab formalism does a mixture of both 
these things by using a relativistic action, the clover action, 
but interpreting it non-relativistically where necessary to 
avoid the largest discretisation errors (El-Khadra \etal 1997). This is the method 
currently being used most extensively for $c$ quarks and results 
will be described in section 3. Another promising method 
is to use anisotropic lattices in which the lattice spacing in the 
time direction is much finer than that in spatial directions. This 
then allows $m_ca_t$ to be small without requiring a huge amount 
of computer power to make a fine grid in 4-dimensions. The price of 
this is a relatively complicated action, however. 

\section{Results}

Recent lattice results which include for the first time 
realistic quark vacuum polarisation effects, have changed the 
landscape of lattice calculations. We will concentrate on 
those results here and the possibilities for the future that 
they engender. 

The new results are based on gluon field configurations generated 
by the MILC collaboration and analysis by the HPQCD and MILC 
collaborations (Davies \etal 2004). They have worked with a highly 
improved gluon action and with improved staggered (asqtad) 
quarks, as described in the previous section. The effect 
of the quark vacuum polarisation is included in the generation 
of the gluon configurations by taking the fourth root of the 
determinant of the improved staggered quark matrix for each 
flavour of quark included in the sea. Three flavours of quarks 
are included in the sea, $u$, $d$ and $s$. This is believed 
to be all that is necessary, since a perturbative analysis 
shows that the effects of $c$ or $b$ quarks in the sea is 
very small (Nobes 2005). The $u$ and $d$ quarks are taken in fact to have 
the same mass (as in all current lattice calculations) because 
this provides for some simplification and should give only a small
error when comparson is made to appropriately isospin-averaged 
experimental quantities. The configurations are then referred 
to as `2+1' flavour unquenched configurations. 

As described above, it is not possible to perform the 
lattice calculations at the physical $u/d$ quark mass
(since we take $m_u=m_d$ this would be $(m_{u,phys}+m_{d,phys})/2$). 
Instead, many different ensembles of configurations are 
made with different values of $m_{u/d}$ and extrapolations 
to find the physical point must be done. These chiral 
extrapolations will be accurate provided that 
we are close enough to the physical point and we are able 
to constrain the functional form of the dependence on 
$m_{u/d}$ with chiral perturbation theory. 

It is convenient, 
when discussing the $u/d$ quark masses used in the 
lattice calculations, to give them in terms of the 
strange quark mass, $m_s$. This is because there is 
no problem (with current computers) in doing lattice 
QCD calculations that include the $s$ quark vacuum 
polarisation. However, it is also clear that, in the 
real world, there is a very visible difference 
between $s$ quarks and $u/d$ quarks. Indeed, if we 
consider chiral perturbation theory as an expansion 
in powers of $x_q = (m_{PS}/(4\pi f_{\pi})^2$ where 
$m_{PS}$ is the mass of the pseudoscalar light meson 
made from the light quarks, then $x_s = 0.33$ and 
$x_{u/d} = 0.03$. We would therefore not expect 
low order chiral perturbation theory to work well 
for light quarks with a mass equal to that of 
the $s$ quark. This is not a problem since we 
do not need chiral perturbation 
theory to reach the $s$ quark mass. However it does 
mean that our $u/d$ quark mass must be well below 
$m_s$ to expect to be able to reach the physical 
point by chiral extrapolation in $m_{u/d}$.  
However, with  a set of ensembles with 
$m_{u/d} < m_s/2$ and using next-to-leading order 
chiral perturbation theory in $m_{u/d}$, it should be possible 
to perform extrapolations with errors at the 
few \% level. This is what has now been done using 
the configurations generated by the MILC collaboration. 
The MILC configurations range in $m_{u/d}$ down to
$m_{u/d} = m_s/8$ which is within a 
factor of three of the real world. 
This is a huge improvement over previous calculations 
which had only two flavours of sea quarks, 
meant to represent $u$ and $d$, but with $m_{u/d} > m_s/2$, 
and it is this improvement that is responsible for 
the quality of the new results. 

In addition to including the effect of quark vacuum 
polarisation with light $u/d$ quarks and $s$ quarks, 
the MILC configurations come in sets with three different 
values of the lattice spacing. In each set the 
bare coupling constant must be adjusted as the 
quark masses are changed to keep the lattice spacing 
approximately the same (as described earlier the lattice 
spacing is determined accurately after the simulation, 
but preliminary results enable this approximate tuning 
of the lattice spacing to be done). Having a set 
of configurations which include the effect of sea quarks 
of different light quark mass but which have the same 
lattice spacing is important to allow the chiral extrapolations 
to be done without confusing the (real) effects of changing $m_{u/d}$ 
with (unphysical) systematic errors from having different values of the 
lattice spacing. The effect of the discretisation errors 
can also be gauged accurately (and an extrapolation to the 
continuum limit be performed if necessary) by having 
several well-spaced values of the lattice spacing.  
Again this has not been possible with previous lattice 
calculations. 
On the MILC configurations the lattice spacing values are approximately 
0.18 fm (supercoarse), 0.12 fm (coarse) and 0.08fm (fine), 
chosen to be appropriately spaced in $a^2$, the expected 
size of the discretisation errors with the improved 
action used. Most of the results described here will come 
from the coarse and fine sets, since the supercoarse set 
has only been made recently. A superfine set is also now 
planned, to give a fourth lattice spacing value. 
In addition another set of ensembles is being made with a 
different bare sea $s$ quark mass. There is no problem in principle 
with working at the correct $s$ mass, but it is hard to tune 
this accurately until after the calculations have been done. 
On the coarse ensemble the sea $s$ mass is in fact 25\% high 
and it is 10\% high on the fine set of ensembles. Results with a 
different $s$ mass will enable a more accurate interpolation to 
the correct $s$ mass.

Whilst singing the praises of the MILC configurations, it should 
also be said that they have a commendably large volume (2.5 fm on a 
spatial side) and a very long extent in the time direction (nearly 
8 fm). There are several hundred configurations in each ensemble
that can be used for calculations. All of these factors lead
to small statistical errors, indeed smaller than for a lot of 
calculations that have been done in the quenched approximation. 
The MILC configurations are publicly available at www.nersc.gov. 

\subsection{Results on the spectrum}

\begin{figure}[ht]
\centering{
\includegraphics[width=80mm, clip, trim=0 0 0 0]{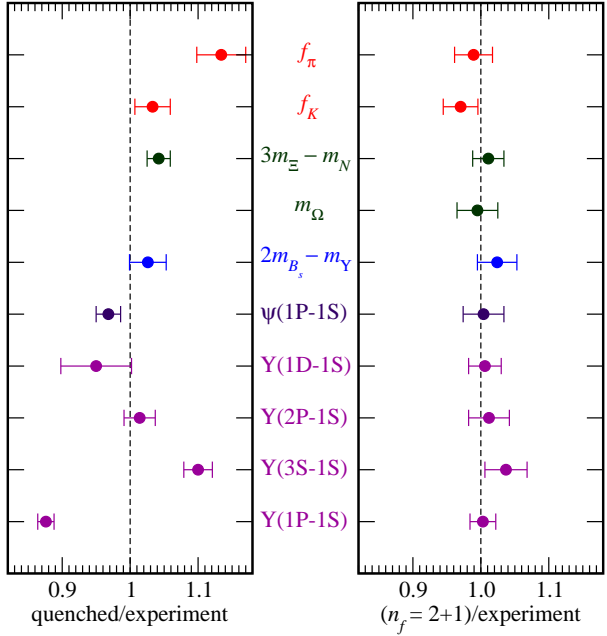}
\caption{Lattice QCD divided by experiment for a range of 
`gold-plated' quantities that cover the full range of QCD physics.
The unquenched calculations on the right show agreement 
with experiment across the board, whereas the quenched 
approximation on the left gives systematic errors of 
$\cal{O}$(10\%). (Davies \etal 2004)}
\label{davies:fig12}}
\end{figure}

\Figure{\ref{davies:fig12}} summarises the new results (Davies \etal 2004). It shows 
the ratio of the lattice calculation to the experimental number 
for a range of gold-plated quantities from the $\pi$ decay constant 
to radial and orbital excitation energies in the $\Upsilon$ system, 
by way of heavy-light meson masses and baryon masses. 
The left-hand panel of the Figure shows results in the old 
quenched approximation and the right-hand panel shows the 
new results including the effect of light quark vacuum 
polarisation. The clear qualitative difference between the 
two panels is that the left-hand panel shows clear disagreement 
with experiment for a number of quantities because the quenched 
approximation is wrong. The right-hand panel shows instead agreement 
with experiment for all the quantities within errors of a few percent. 

To make \Figure{\ref{davies:fig12}} the parameters of QCD first 
had to be fixed. The lattice spacing was determined from 
the radial excitation energy in the $\Upsilon$ system 
($ 2S-1S \equiv M_{\Upsilon^{\prime}}-M_{\Upsilon}$). The $u/d$ quark mass was 
fixed by finding where the $\pi$ mass would be correct 
by extrapolation using chiral perturbation theory. The $s$ quark 
mass was fixed by determining where the $K$ meson mass was 
correct. The $c$ quark mass was fixed using the $D_s$ and 
the $b$ quark mass was fixed by getting the $\Upsilon$ mass 
correct. All of the hadron masses used to determine the 
parameters are `gold-plated' {\textit{ie}} they have very small 
decay widths and are well below strong decay thresholds. 
This means that they are well-defined experimentally and 
theoretically and should be accurately calculable in lattice 
QCD. Using them to fix parameters will not then introduce 
unnecessary additional systematic errors into lattice 
results for other quantities. This is an important issue 
when lattice QCD is to be used as a precision calculational 
tool. 

\begin{figure}[ht]
\centering{
\rotatebox{270}{\includegraphics[width=63mm, clip, trim=10 5 0 0]{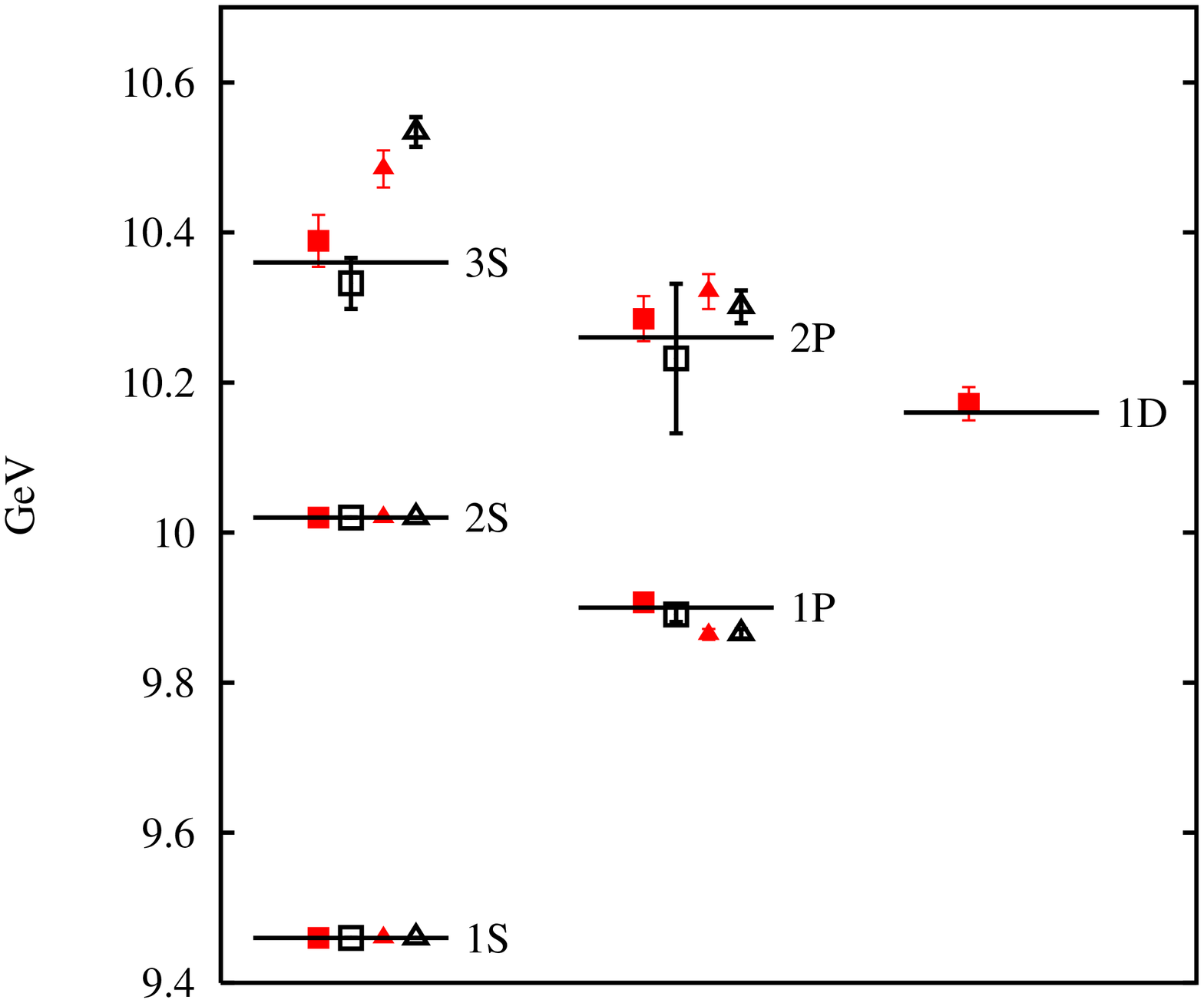}}\\
\rotatebox{270}{\includegraphics[width=70mm, clip, trim=0 0 0 90]{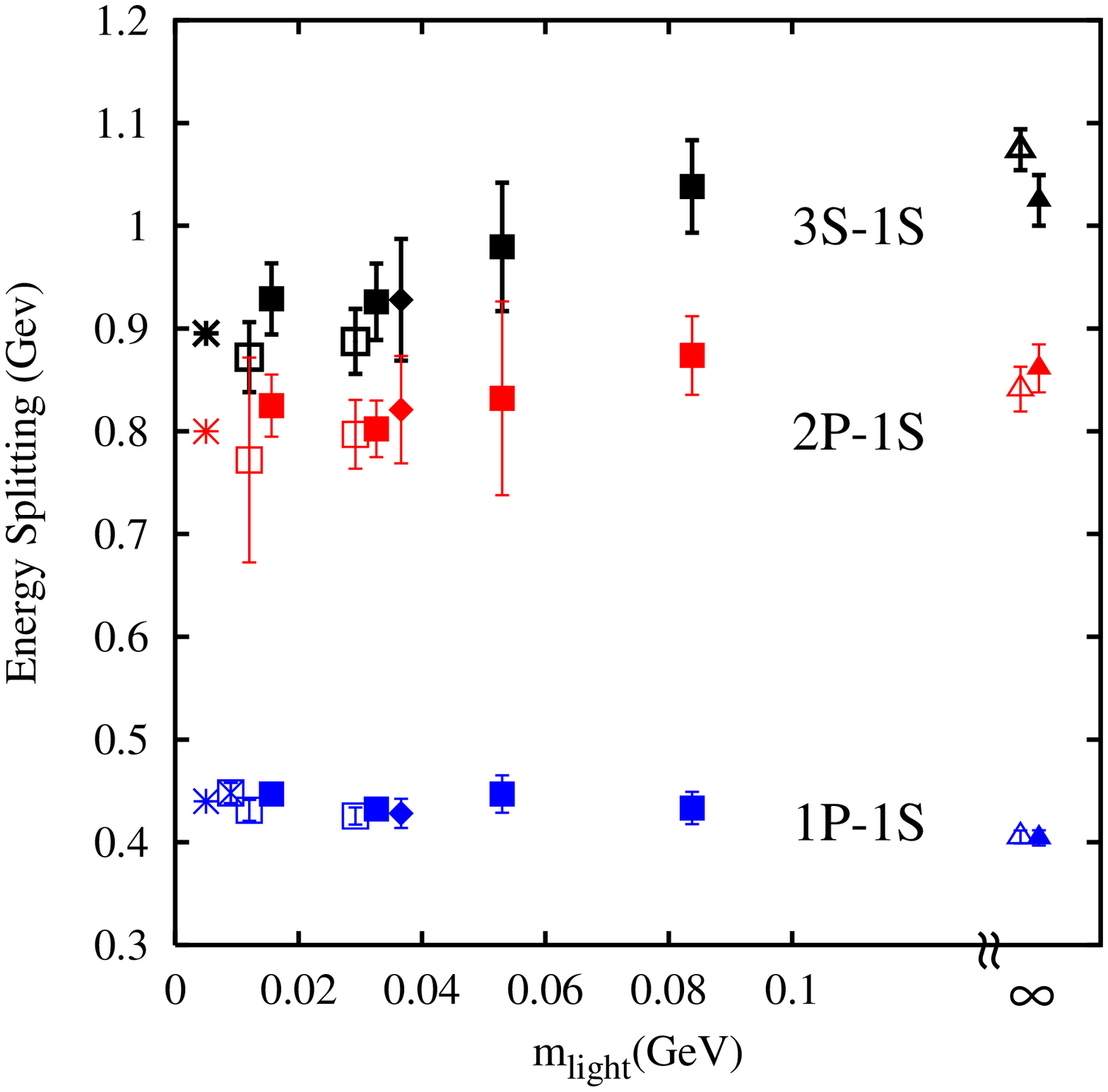}}
\caption{Radial and orbital splittings in the $\Upsilon$ spectrum 
from lattice QCD in the quenched approximation and including 
a realistic light quark vacuum  polarisation. In these plots the 
$b$ quark mass was fixed from the $\Upsilon$ mass and the lattice 
spacing from the splitting between the $\Upsilon^{\prime}$ and the 
$\Upsilon$. Neither of these masses is predicted. 
(Top) The spectrum of $S$, $P$ and $D$ levels in the $\Upsilon$ 
system obtained from coarse (filled red triangles) and fine 
(open black triangles) quenched lattice calculations and from 
coarse (filled red squares) and fine (open black squares) unquenched 
calculations. Experimental results are shown as lines. 
(Bottom) Results for different splittings as a function of light $u/d$ 
quark mass. The leftmost points, at lightest $u/d$ quark mass, 
are the ones included in the top plot for the unquenched results. (Gray \etal 2003)}
\label{davies:fig13}}
\end{figure}

Having fixed the parameters, we can then focus on other 
gold-plated masses and decay constants and \Figure{\ref{davies:fig12}}
shows the predictions for these other quantities. The fact 
that the right-hand panel demonstrates agreement with experiment for 
all the quantities shown is an indication that the parameters 
of QCD are unique. Instead of using the hadron masses of 
the previous paragraph to fix the parameters we could have 
used appropriate quantities from \Figure{\ref{davies:fig12}} and 
we would have obtained the same answer (and would then have been 
able to predict $m_{\pi}$, $m_K$ etc). The left-hand panel 
shows that this is not true in the quenched approximation. 
The results there show disagreement with experiment and it 
is clear, for example, that if we had used the orbital 
excitation energy in the $\Upsilon$ system  ($1P-1S$) 
to fix the lattice spacing we would have obtained an answer 
10\% different. The quenched approximation is then internally 
inconsistent since the parameters depend on the hadrons 
used to fix them.  The basic reason is that, as the light 
quark vacuum polarisation is missing, the strong coupling constant 
runs incorrectly between different momentum scales. Therefore 
hadrons which are sensitive to different momentum scales 
cannot simultaneously agree with experiment.  

In \Figure{\ref{davies:fig13}} more details are shown of the 
lattice results for the radial 
and orbital energy levels in the $\Upsilon$ system 
in the old quenched approximation and now using 
the MILC configurations with their inclusion of a 
realistic light quark vacuum polarisation (Gray \etal 2003). 
Our physical understanding of the $\Upsilon$ system is very good 
and there are a lot of gold-plated states below decay threshold 
so it is a very valuable system for lattice QCD tests and for 
determining the lattice spacing. We use the 
standard lattice NRQCD effective theory, described above, for the valence 
$b$ quarks, accurate through $v^4$ where $v$ is the velocity 
of the $b$ quark in its bound state. This means that 
spin-independent splittings, such as radial and orbital 
excitations, are simulated through next-to-leading-order 
and should be accurate to about 1\%. The test of QCD using 
these splittings is then a very accurate one. 
We do not expect the $\Upsilon$ system to be very sensitive 
to the masses of the light quarks included in the quark 
vacuum polarisation, only to their number. The momentum 
transfer inside an $\Upsilon$ is larger than any of the 
$u$, $d$, or $s$ masses and so we expect these splittings 
simply to `count' the presence of the light quarks. 
This lack of variation with light quark mass is 
evident in \Figure{\ref{davies:fig13}}. 

\begin{figure}[ht]
\centering{
\rotatebox{270}{\includegraphics[width=60mm, clip, trim=0 10 0 100]{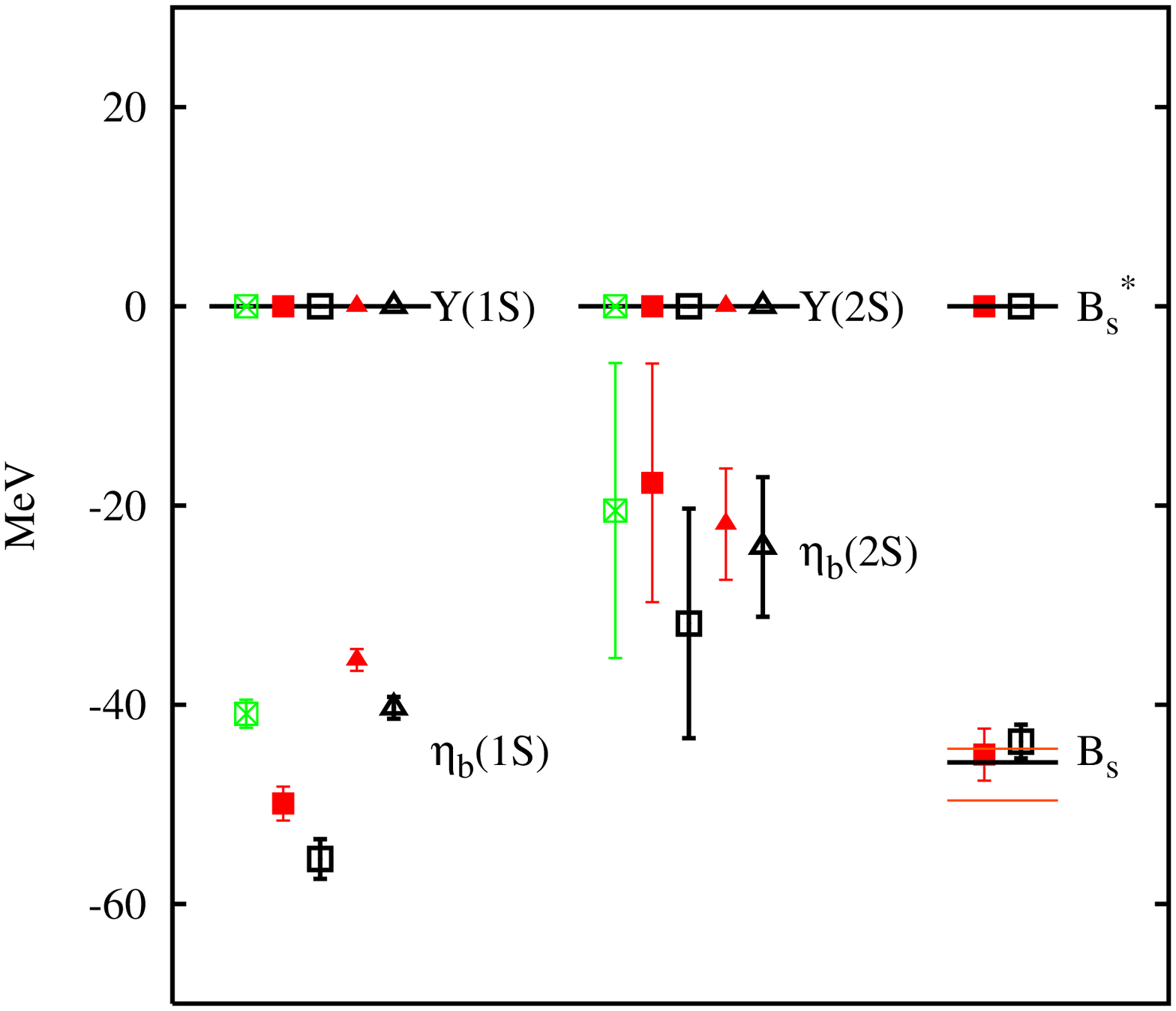}}\\
\rotatebox{270}{\includegraphics[width=61mm, clip, trim=0 10 0 100]{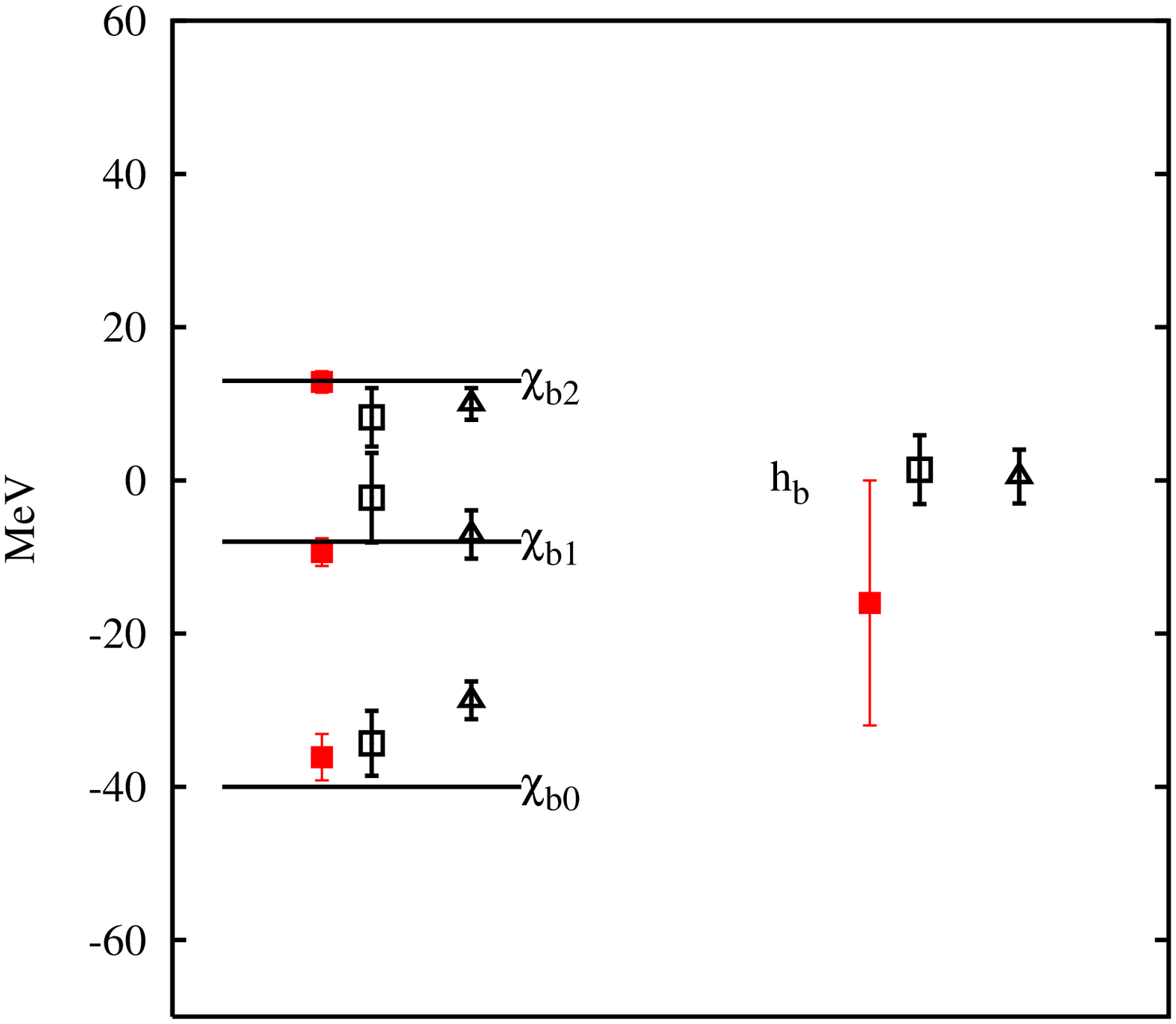}}
\caption{Fine structure in the $\Upsilon$ system. Triangles 
show quenched results (closed red=coarse, open black=fine) and squares 
show unquenched results (crossed green=supercoarse, closed red=coarse, open black=fine). 
(Top) $S$-wave splittings between $\Upsilon$ and $\eta_b$ and 
between $\Upsilon^{\prime}$ and $\eta_b^{\prime}$. The same 
NRQCD action is used to determine the splitting between 
the $B_s^*$ and the $B_s$ and this is shown on the right. There 
the experimental situation is summarised by the lines - faint 
lines show limits for the experimental $B_s$ results, the darker 
line shows the more precise result for the $B$, which is expected 
to be very close to the result for the $B_s$.  
(Bottom) $P$-wave splittings between the $^3P_{0,1,2}$ $\chi_b$ states, compared to 
experiment. The results on the right show the lattice prediction 
for the unseen $^1P_1$ $h_b$ state. (Gray \etal 2003)}
\label{davies:fig14}}
\end{figure}

\Figure{\ref{davies:fig14}} shows the fine structure in the 
$\Upsilon$ spectrum. Since the fine structure is determined 
at leading order by spin-dependent terms that appear 
first at $\cal{O}$$(v^4)$ this has significant systematic errors 
from missing higher order terms. There are also 
systematic errors arising from missing radiative corrections 
to the leading $v^4$ terms. In fact, encouraging agreement 
with experiment is obtained and we are able to predict the 
ground-state hyperfine splitting between the 
$\Upsilon$ and the unseen $\eta_b$ to be 60(15) MeV.  
The results for the fine structure can be improved by improving 
the NRQCD action and this will be done in the near future. 
For comparison to the $\Upsilon$ splittings, \Figure{\ref{davies:fig14}} 
also shows the splitting between corresponding mesons made 
of one $b$ quark and one $s$ quark {\textit{ie}} the $B_s$ heavy-light 
system. This is calculated on the lattice using the same 
NRQCD action of the valence $b$ quarks and the improved staggered 
action (as used in the quark vacuum polarisation) for 
the light quark. Once the $b$ quark mass and lattice spacing 
have been fixed from the $\Upsilon$ system the heavy-light 
spectrum is entirely predicted by lattice QCD and provides 
another useful test against experiment. 

\begin{figure}[ht]
\centering{
\includegraphics[width=62mm, clip, trim=0 40 0 110]{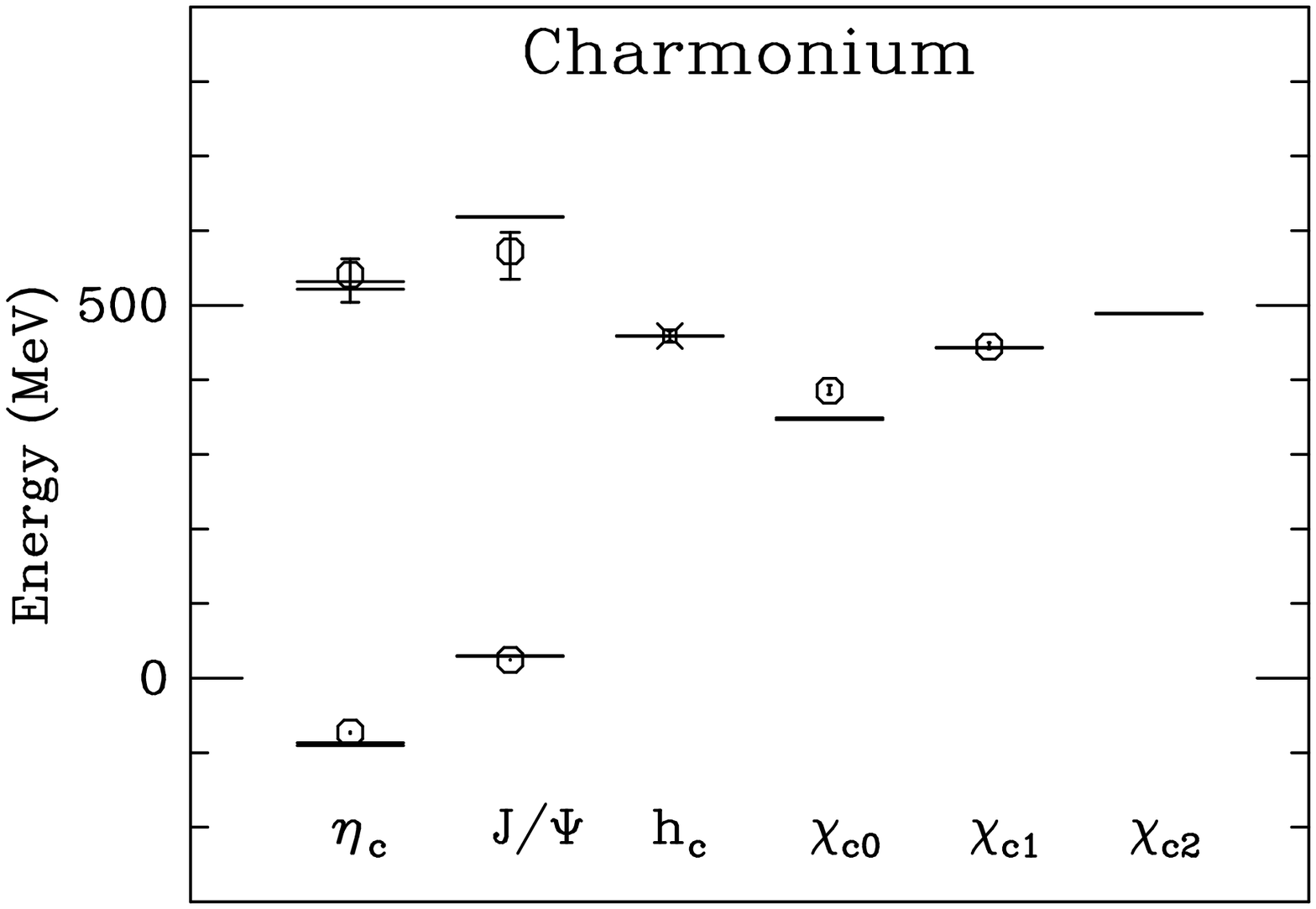}
\includegraphics[width=60mm, clip, trim=0 40 20 110]{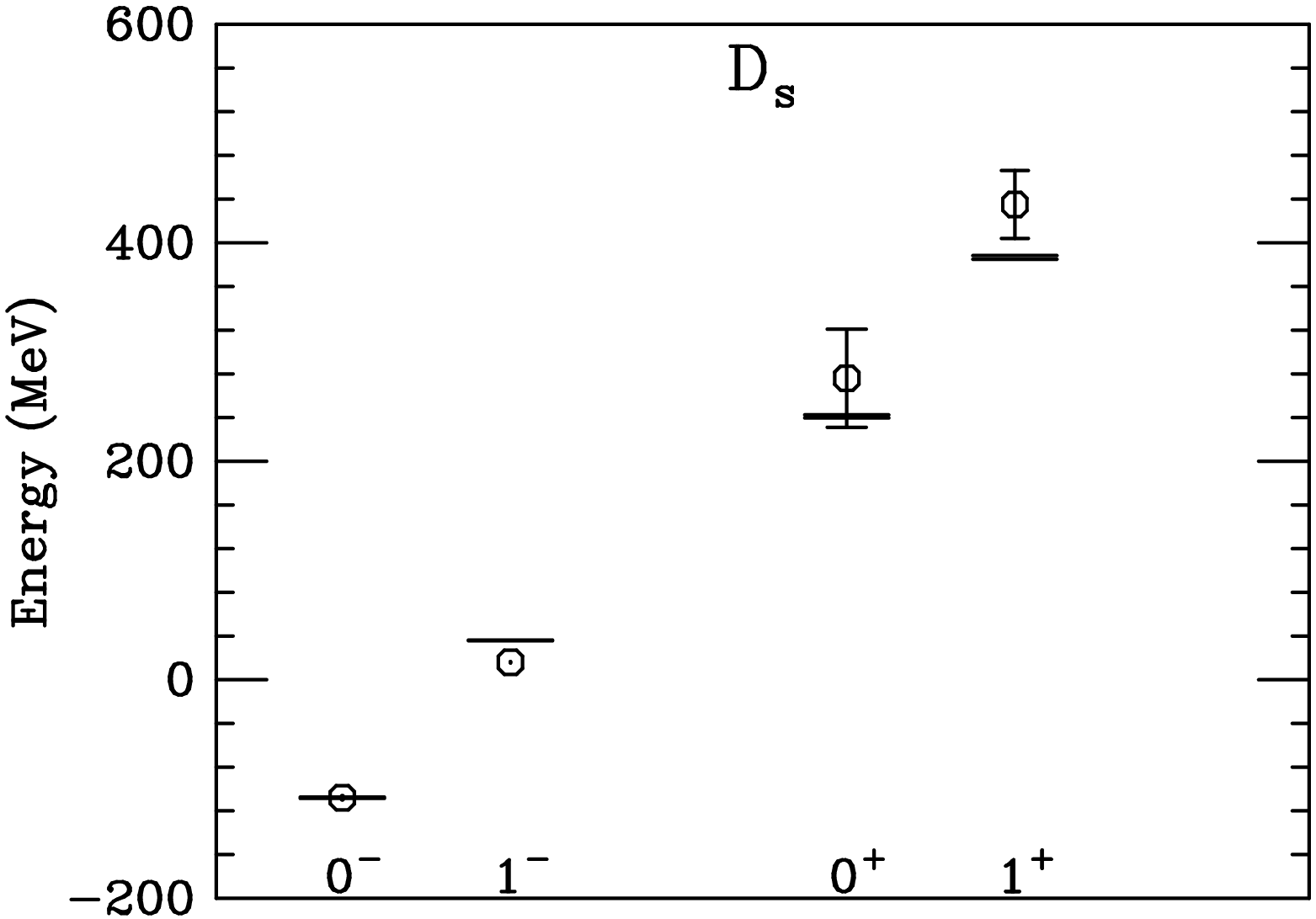}
\caption{Spectrum of mesons containing charm quarks 
from lattice QCD including light quark vacuum polarisation. 
These results are from the coarse unquenched MILC configurations 
using the $\Upsilon$ system to determine the lattice spacing and 
the $D_s$ meson mass to fix the $c$ quark mass. The Fermilab 
action was used for the $c$ quark. 
(Left) The charmonium system. 
(Right) The $D_s$ system. 
(di Pierro \etal 2004)}
\label{davies:fig15}}
\end{figure}

The charm quark is somewhat too light for good results using the 
NRQCD action above. The results shown in \Figure{\ref{davies:fig15}}
use the Fermilab action described earlier which aims to 
interpolate between light relativistic quarks and the heavy 
nonrelativistic limit (di Pierro \etal 2004). It should be borne in mind that there are 
fewer gold-plated states in the $c\overline{c}$ spectrum compared 
to the $b\overline{b}$ spectrum described above. In fact even 
the $\psi^{\prime}$ is rather close to decay threshold to be sure that 
its mass is not affected by coupling to virtual decay modes which 
will have a distorted momentum spectrum on a finite lattice volume. 
The right-hand plot of \Figure{\ref{davies:fig15}} shows the $D_s$ 
spectrum. The $D_s$ mass was used to fix the $c$ quark mass, so 
that is not predicted. The two newly discovered $0^+$ and $1^+$ 
states are shown. The simplest explanation for these states is 
that they are $P$-wave states, narrow because their masses are 
too low for the Zweig-allowed decay to $D, K$. They do decay to 
$D_s, \pi$, however, and are therefore not gold-plated. 
We expect systematic errors in a lattice QCD calculation
of their masses, even when light quark vacuum polarisation is included. 
The current lattice results, shown in \Figure{\ref{davies:fig15}}, are 
consistent with this picture and certainly do not require any 
more exotic explanation of the meson internal structure. 

\begin{figure}[ht]
\centering{
\includegraphics[height=42mm, clip, trim=0 10 10 5]{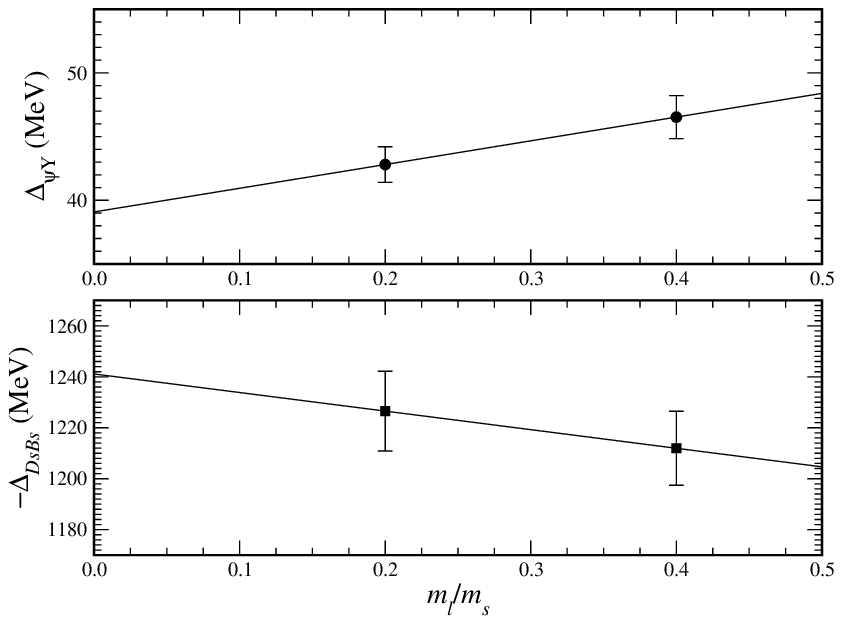}
\includegraphics[height=40mm, clip, trim=0 1 8 130]{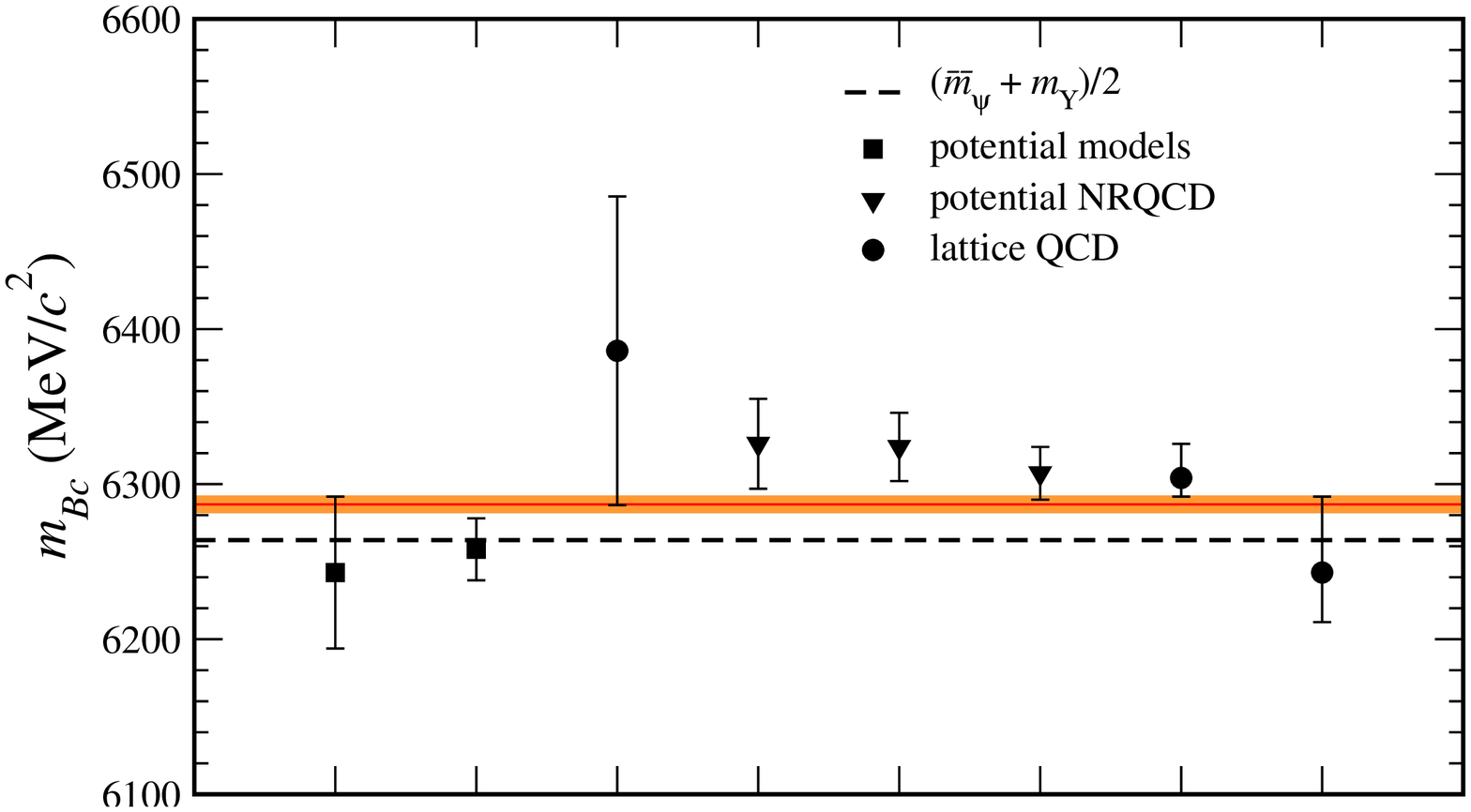}
\caption{(Left) The differences between the $B_c$ mass and, 
above, the average of $\Upsilon$ and $\psi$ masses and, 
below, the $B_s$ and $D_s$ masses, plotted against the 
light quark mass included in the quark vacuum polarisation 
in the MILC unquenched configurations. 
(Right) Predictions for the mass of the $B_c$ from a variety of 
methods. In the centre is the old quenched lattice QCD result 
and on the right the two new lattice QCD predictions. The dashed 
line shows where the $B_c$ mass would be equal to the average 
of $\Upsilon$ and $\psi$ masses and the broader coloured line is the 
new experimental result. (Allison \etal 2005, Acosta \etal 2005)}
\label{davies:fig16}}
\end{figure}

With a good action for $b$ quarks and a good action for $c$ quarks
it is possible to predict the mass of the $B_c$ meson. The results 
are shown in \Figure{\ref{davies:fig16}}, (Allison \etal 2005) compared to the very 
recent experimental results (Acosta \etal 2005). The quantity that is calculated on 
the lattice is the mass difference between the $B_c$ and 
a combination of masses of other mesons containing a $b$ quark 
and a $c$ quark, enabling some of the systematic errors to 
cancel. There are two combinations available on the lattice. 
The one that gives the most accurate result we believe is 
to calculate the difference between the $B_c$ mass and that 
of the average of the $\Upsilon$ and $\psi$. This is 
called $\Delta_{\psi\Upsilon}$. It turns out to be very small, 
although non-zero. The difference between the 
$B_c$ and the sum of the masses of the $D_s$ and $B_s$, $\Delta_{D_sB_s}$, is  
also useful but less accurate. Both splittings are shown in 
\Figure{\ref{davies:fig16}} as a function of the light 
quark mass included in the quark vacuum polarisation on the 
MILC unquenched configurations. After a study of systematic 
errors from fixing the quark masses and lattice spacing we 
are able to give a final result for the $B_c$ mass of 
6304(20) MeV (Allison \etal 2005). The error is much better than the 100 MeV obtained 
for previous calculations in the quenched approximation. There 
100 MeV arose directly from the impossibility of consistently 
determining the parameters of the theory in the quenched 
approximation, a problem that disappears in the new 
unquenched results. The very recent experimental result from CDF is 
6287(5) MeV, showing impressive agreement with the unquenched 
lattice calculation (Acosta \etal 2005). 

\begin{figure}[ht]
\centering{
\includegraphics[height=55mm, clip, trim=0 0 0 2]{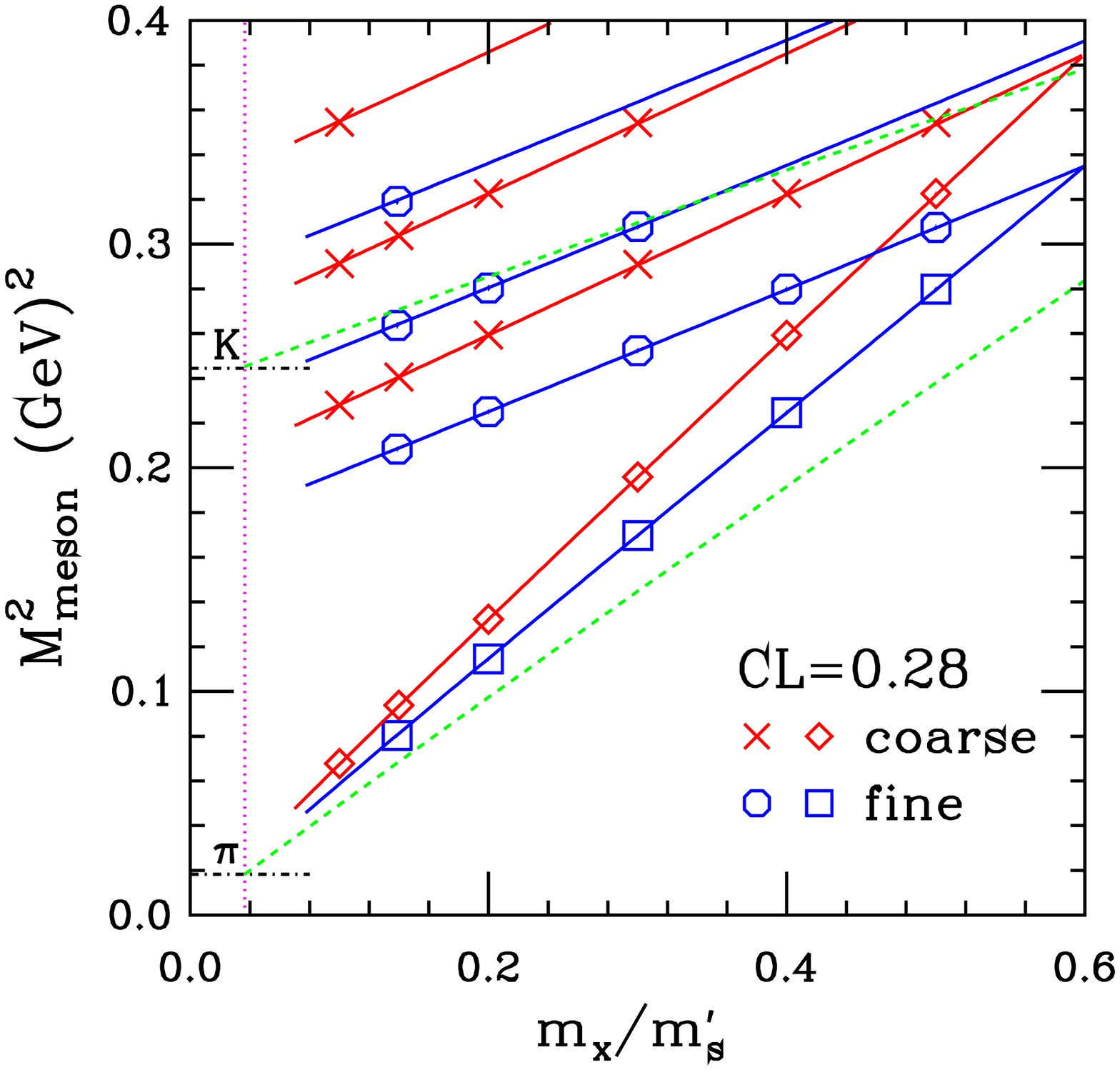}
\includegraphics[height=55mm, clip, trim=0 10 0 10]{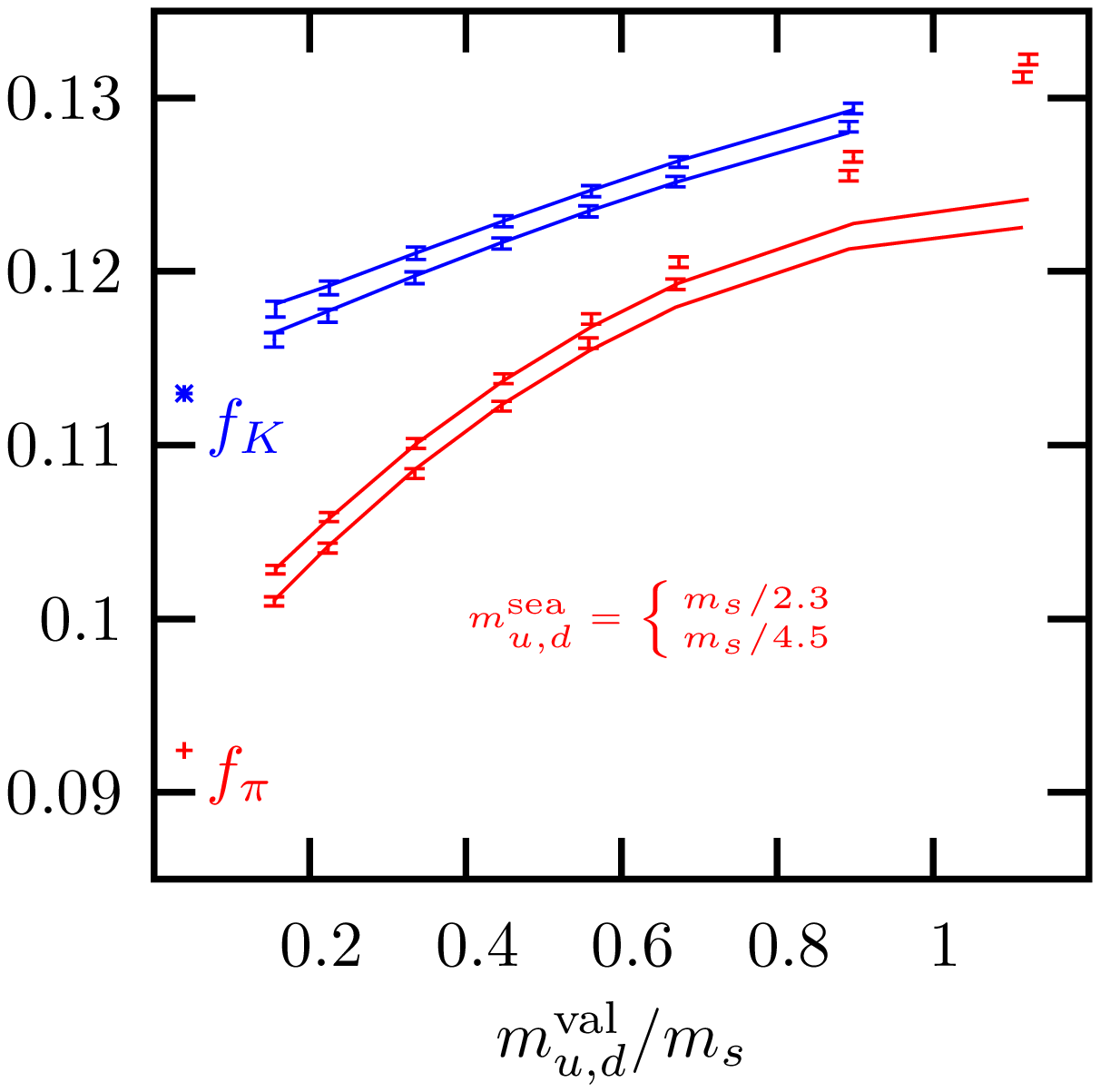}
\caption{(Left) Chiral fits and extrapolations for light meson ($\pi$ and 
$K$) meson masses from the MILC configurations. The square of the 
meson mass is plotted against the light ($u/d$) quark mass used in the 
quark vacuum polarisation. This is denoted $m_X$ and given in units of 
the $s$ quark mass used in the quark vacuum polarisation (denoted here $m_s^{\prime}$). 
Several different values of valence $s$ quark mass were used so there are 
several different sets of results for the $K$ mass. 
The lattice results are the points and lines are high-order chiral perturbation 
theory fits. At small light quark mass these are indistinguishable from 
the lowest order straight line.   
(Right) Chiral fits for $f_{\pi}$ and $f_K$, the $\pi$ and $K$ decay constants. 
The points are lattice calculations on the fine unquenched MILC configurations, 
for two different values of the light quark mass in the quark vacuum polarisation 
($m^{sea}_{u/d}$). 
The results are plotted against the valence $u/d$ quark mass in units
of the valence $s$ quark mass.
Note that only points with $m_{u/d} < m_s/2$ (for both valence and sea) were used 
in the chiral fits, shown as lines. The fits 
were then extrapolated back to compare to the lattice results with 
$m^{valence}_{u/d} > m_s/2$. (Aubin \etal 2004a, 2004b, Lepage and Davies 2004)}
\label{davies:fig17}}
\end{figure}

Results for light mesons are equally good on the MILC configurations (Aubin \etal 2004a, 2004b, 
Bernard 2001). 
\Figure{\ref{davies:fig17}} shows results for the light meson 
masses and decay constants plotted against light quark 
mass. The left-hand plot shows the fits to chiral perturbation 
theory for the squared masses of the $\pi$ and $K$, which must then 
be extrapolated (for the $\pi$) or interpolated (for the $K$) to 
find the quark masses which correspond to $u/d$ (bearing in 
mind that the $u$ and $d$ quark masses have been taken to be the 
same) and $s$. Once this has 
been done, predictions for other light meson quantities can be made. 
Two very clean quantities to calculate are the light meson decay constants, 
$f_{\pi}$ or $f_K$. 
The decay constant is obtained from the matrix element of the temporal axial current 
between the pseudoscalar light meson 
and the QCD vacuum and it is related, for the charged $\pi$ and $K$, 
to the rate of purely leptonic decay. This rate is well-known experimentally 
for these mesons
and can be calculated accurately on the lattice. This is because, unlike 
the matrix elements we will discuss below, the axial current needs 
no renormalisation for improved staggered quarks. The right-hand plot 
of \Figure{\ref{davies:fig17}} shows the chiral extrapolations for 
the $\pi$ and $K$ decay constants for the fine MILC lattices. Only 
points for which the $u/d$ quark mass is less than half the 
$s$ quark mass are used in the fits to chiral perturbation theory. 
The curves, however, show what happens when these fits are 
extrapolated to higher $u/d$ quark masses. For $f_{\pi}$ it is 
clear that the heavier $u/d$ quark are not on the same chiral 
perturbation theory curve as the lighter ones {\textit{ie}} using heavier 
$u/d$ masses would have distorted the chiral fit and given the 
wrong result in the chiral limit where the $u/d$ quark mass 
takes the correct value to give the experimental $\pi$ meson mass. 
This should be borne in mind when looking at earlier lattice results 
that have only $m_{u/d} > m_s/2$.   

\begin{figure}[ht]
\centering{
\includegraphics[height=55mm, clip, trim=0 0 0 0]{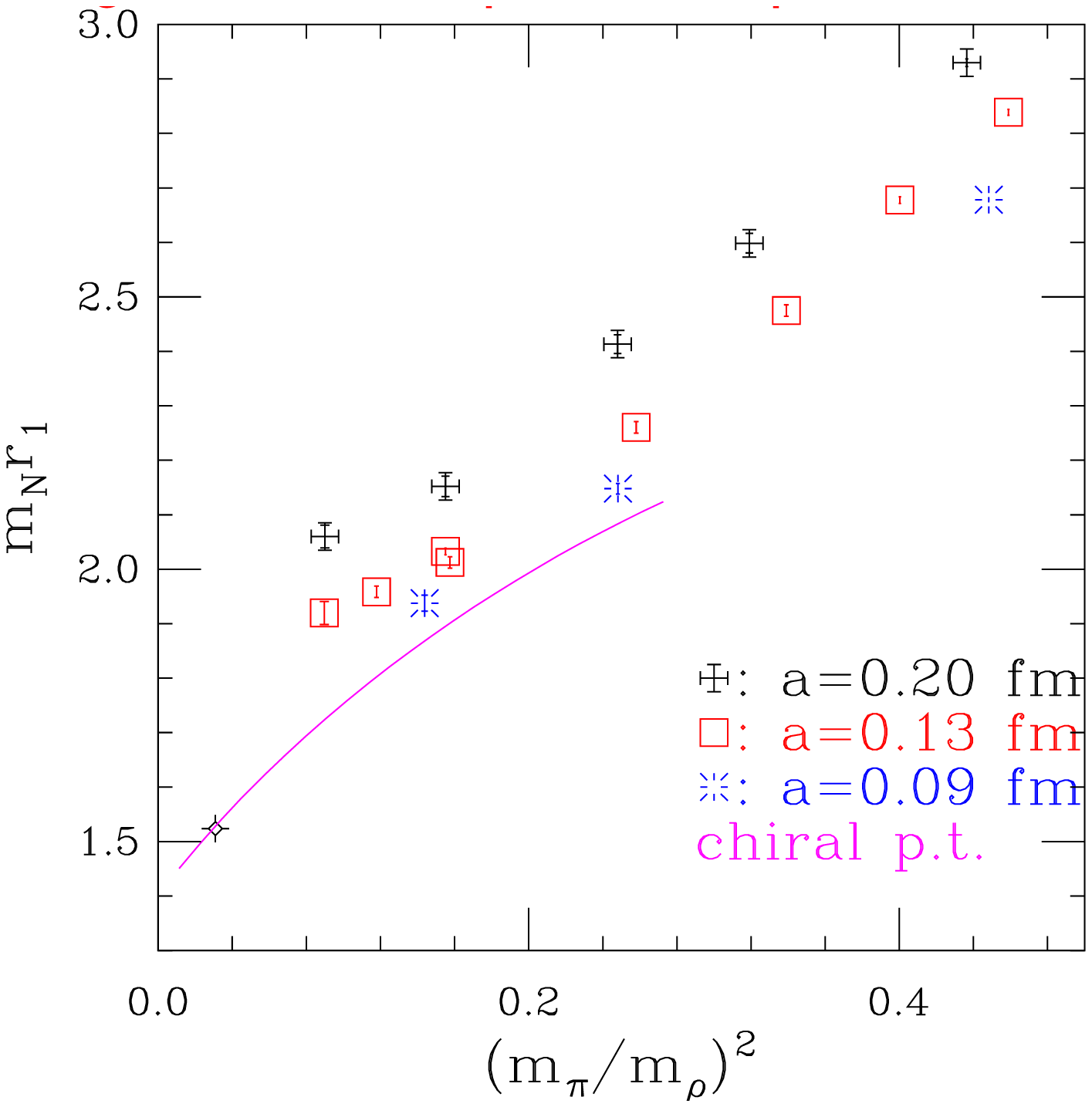}
\includegraphics[height=55mm, clip, trim=0 0 0 0]{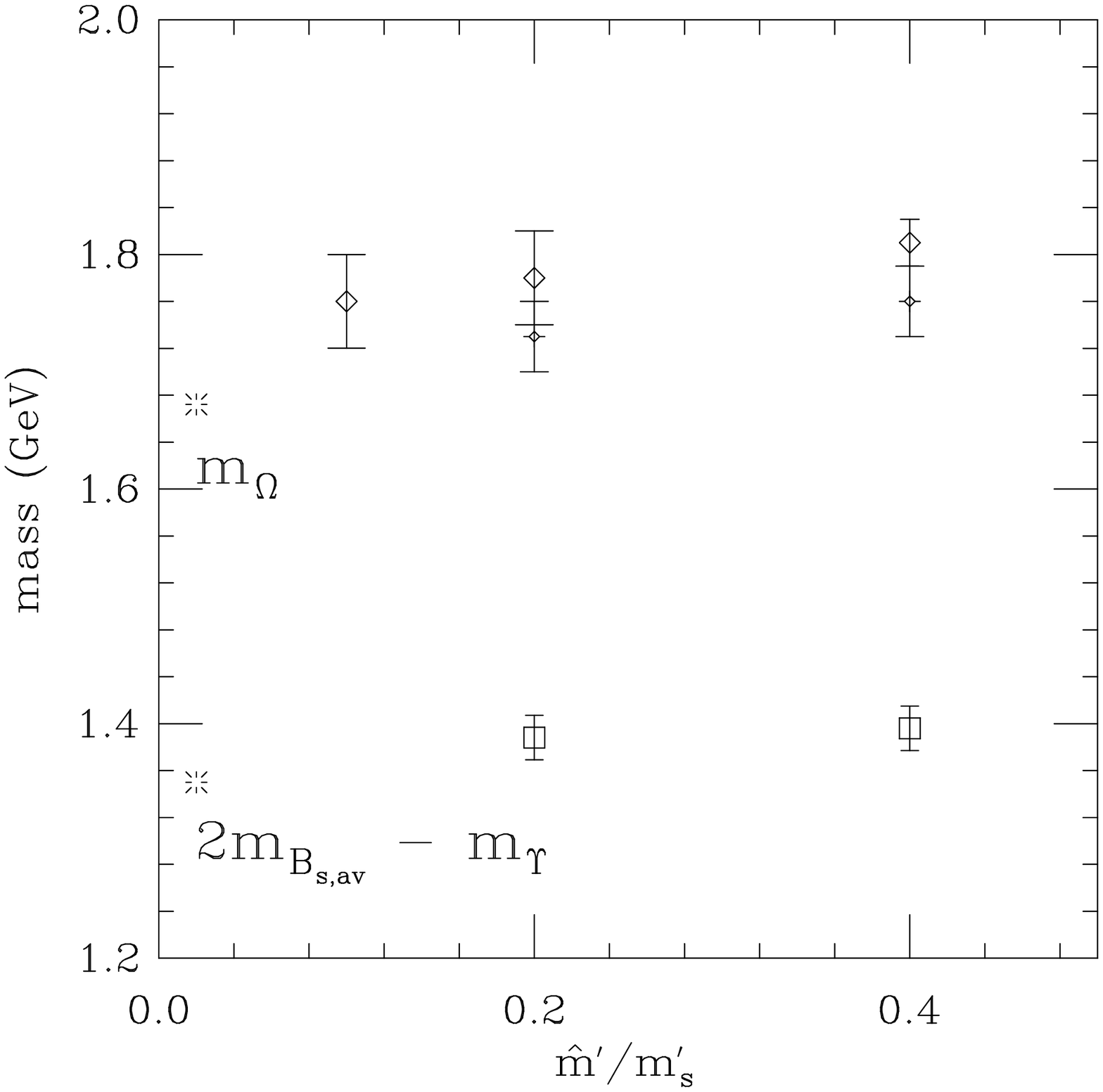}
\caption{(Left) The nucleon mass in units of the $r_1$ parameter from 
the unquenched MILC configurations at three different values of 
the lattice spacing and for several values of the light quark 
mass, denoted on the $x$ axis by the ratio of $\pi$ to $\rho$ masses 
made of the light quarks. The line shows continuum chiral perturbation 
theory for this quantity (Aubin \etal 2004a). The fancy diamond 
gives the experimental point in these units.  
(Right) The $\Omega$ baryon mass as a function of sea light quark 
mass $m^{\prime}$ divided by sea $s$ quark mass. The $\Omega$ mass 
is calculated for a valence $s$ quark mass which is the correct one, fixed 
from the $K$ (Toussaint and Davies 2005). The burst gives the 
experimental point. Also on this plot is the splitting between the 
spin average of the $B_s$ and $B_s^*$ and the $\Upsilon$, a quantity 
sensitive to the $s$ quark mass, but not the $b$ quark mass. Again it 
shows good agreement with experiment denoted by a burst (Aubin \etal 2004c).}
\label{davies:fig18}}
\end{figure}

Light baryons are also gold-plated particles and calculating their 
masses provides a further opportunity for predictions from lattice 
QCD since all the QCD parameters have been fixed from the meson sector. 
Baryons are harder to work with on the lattice and the masses are 
not as precise. The nucleon also requires, for example, a more 
complicated chiral fit and this has not yet been done. 
\Figure{\ref{davies:fig18}} shows the nucleon results on the 
super-coarse, coarse and fine MILC unquenched configurations (Aubin \etal 2004a). 
There are signs that the results change slightly with lattice 
spacing {\textit{ie}} there is a visible discretisation error. This needs 
to be studied further. The line shows the low order chiral perturbation 
theory result and the lattice results do seem to be heading towards 
that on the finer lattices. The right-hand plot shows results 
for the $\Omega$ baryon made of three valence $s$ quarks (Toussaint and Davies 2005). This 
baryon is not very dependent on the $u/d$ quark mass used 
in the quark vacuum polarisation so chiral extrapolations are 
much simpler. Its mass is sensitive to the $s$ quark mass, however, so 
the fact that agreement with the experimental result is obtained with the 
$s$ quark mass fixed from the $K$ is another strong confirmation 
that lattice QCD is internally consistent once realistic 
quark vacuum polarisation is included. 

It is important to realise that accurate lattice QCD results are not 
going to be obtainable in the near future for every hadronic quantity 
of interest. What these results show is that `gold-plated' 
quantities should be now be calculable. Unstable hadrons, or even 
those within 100 MeV or so of Zweig-allowed decay thresholds, have 
strong coupling to their real or virtual decay channel and this is 
not correctly simulated on the lattice volumes currently being used. 
{\textit{eg}} the smallest non-zero momentum on typical current lattices 
exceeds 400 MeV. This could significantly distort the decay channel 
contribution to the hadron mass. 
Much larger volumes will then be necessary to handle these hadrons. 
Unfortunately the list of non-gold-plated hadrons is a long one - it 
includes the $\rho$, $\phi$, $D^*$, $\Delta$, $N^*$, glueballs, hybrids {\textit{etc.}} 
Some of these may be more accurately calculable than others and 
qualitative results may also be useful. These points must be borne in 
mind, however, when making quantitative comparison between lattice 
QCD and experiment.  

\subsection{Determination of the parameters of QCD}

Lattice QCD calculations are an excellent way to determine the 
parameters of QCD, masses and coupling constant, because
the method is a very direct one. 

\begin{figure}[ht]
\centering{
\rotatebox{270}{\includegraphics[width=70mm, clip, trim=0 0 0 0]{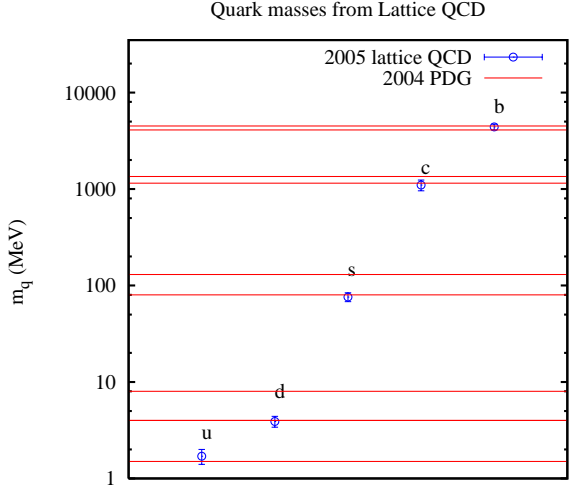}}
\caption{Quark masses in the $\overline{MS}$ scheme at a 
relevant scale (2 GeV for $u$, $d$ and $s$ and their own 
mass for $c$ and $b$) as determined from lattice QCD 
using the unquenched MILC configurations (Aubin \etal 2004c, 2004b, 
Gray \etal 2003, Nobes \etal 2005). The lines 
give the range of the current values quoted in the 
Particle Data Tables (PDG 2004).}
\label{davies:fig19}}
\end{figure}

For the quark masses we 
directly determine the bare quark masses in the lattice QCD 
Lagrangian required to give the correct answer for a given 
hadron mass. With the inclusion of light quark vacuum polarisation 
we have seen above that this can be done both accurately and 
unambiguously. Masses are more conventionally quoted in the 
continuum $\overline{MS}$ renormalisation scheme rather than 
the lattice scheme. To convert from the lattice scheme to 
$\overline{MS}$ requires the calculation of a finite 
renormalisation factor to take into account the 
gluon radiation with momenta larger than $\pi/a$ that 
does not exist on the lattice. The renormalisation factor 
can be calculated in perturbation theory since it involves 
large momenta, and it then appears as a power series in 
$\alpha_s$. This calculation has been done to $\cal{O}$$(\alpha_s)$ 
for the light quark 
masses for the improved staggered quark action (Hein \etal 2003). The $u/d$ quark mass
was fixed from $m_{\pi}$ and the $s$ quark mass from $m_K$ as 
described above. They were then converted to the $\overline{MS}$ 
scheme using the renormalisation factor. This gave, quoting 
the masses at the conventional scale:
\begin{eqnarray}
m_s^{\overline{MS}}(2 GeV) &=& 76(8) MeV, \nonumber \\
m_{u/d}^{\overline{MS}}(2 GeV) &=& 2.8(3) MeV,
\end{eqnarray}
where $m_{u/d}$ is the average of $u$ and $d$ masses (Aubin \etal 2004c). The main 
source of error from the lattice calculation is from unknown 
higher order terms in the perturbative renormalisation 
factor to convert to $\overline{MS}$. A two-loop calculation 
will be available shortly and this will reduce the error to a 
few \%. This is a huge improvement over previous determinations 
of the masses from {\textit{eg}} QCD sum rules. \Figure{\ref{davies:fig19}} shows 
the results for all 5 quark masses ($u$ and $d$ are determined 
separately by matching to different combinations of charged and 
neutral pions and kaons) obtained on the MILC configurations and using  
1-loop matching to convert to the conventional $\overline{MS}$ scheme, 
compared to the results from the 2004 Particle Data Tables (PDG 2004). 
It is quite clear that the lattice will take the lead in providing 
accurate quark masses now. 

\begin{figure}[ht]
\centering{
\includegraphics[height=50mm, clip, trim=0 0 0 0]{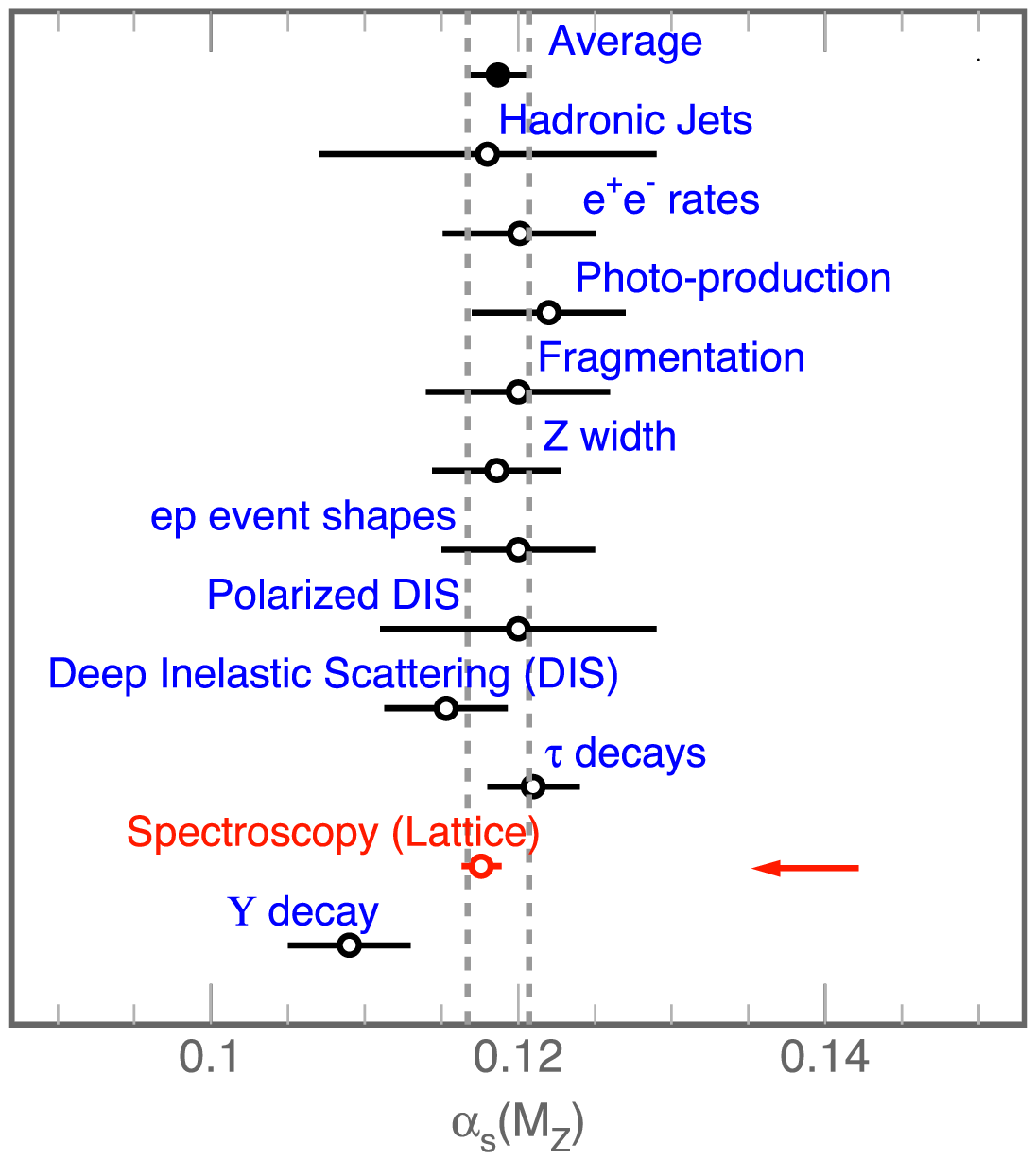}
\includegraphics[height=50mm, clip, trim=0 0 0 0]{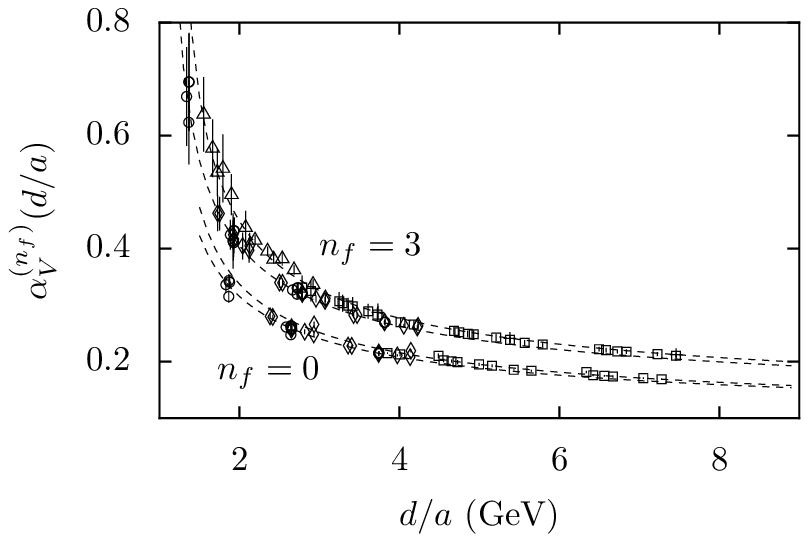}
\caption{(Left) A comparison of the new lattice determination of 
$\alpha_s$, 0.1170(12) with the results from other determinations in the 
Particle Data Tables (PDG 2004).
(Right) The determination of $\alpha_s$ from the lattice in the 
$V$ scheme on quenched ($n_f=0$) and unquenched ($n_f=3$) gluon
configurations at various energy scales, $d/a$ for a variety of Wilson 
loop operators. The dashed curves show the expected running of 
$\alpha_s$ in the two cases from 
QCD. (Mason \etal 2005)}
\label{davies:fig20}}
\end{figure}

The strong coupling constant, $\alpha_s$, is also 
well-determined on the lattice (Mason \etal 2005). The determination 
of $\alpha_s$ proceeds by the calculation in perturbation 
theory to high order of some lattice operator (Trottier 2004). Recently 
calculations up to and including terms in $\alpha_s^3$ 
became available for a lot of different Wilson loops 
and combinations of them. These could then be readily 
`measured' ({\textit{ie}} calculated) on the MILC configurations. 
>From each non-perturbative lattice calculation compared 
to perturbation theory a value for $\alpha_s$ is 
extracted at some momentum scale in lattice units. 
>From the determination of the lattice spacing, this 
scale can then be converted to a physical scale in 
GeV and $\alpha_s$ evolved to different scales use 
the QCD $\beta$ function. Because results at three 
different values of the lattice spacing are available, 
it is possible to do a consistency check for the 
determination of $\alpha_s$ at a given physical scale 
from three different determinations of a particular 
Wilson loop at three different lattice scales. This 
allows estimates of fourth order terms in the 
perturbation theory, which puts this calculation into 
a new regime of accuracy for $\alpha_s$ determinations. 
Altogether 28 different loops and loop combinations 
were studied. The $\alpha_s$ determined was converted 
to the $\overline{MS}$ scheme and run to the 
scale of $M_Z$ since again this is the conventional 
comparison point. The final answer obtained is 0.1170(12) 
which compares very well with other determinations quoted 
in the Particle Data Tables, see \Figure{\ref{davies:fig20}}.

Another interesting point is that $\alpha_s$ is well 
able to distinguish between quenched and unquenched 
configurations. \Figure{\ref{davies:fig20}} shows corresponding 
determinations of $\alpha_s$ from MILC quenched and 
unquenched configurations for different Wilson loops. 
The relevant momentum for the $\alpha_s$ determination 
varies from loop to loop and also depends on whether the 
result comes from the super-coarse, coarse or fine lattices. 
This enables a comparison of $\alpha_s$ values with the 
expected curve for the running. The agreement is very good 
and shows both that the quenched ($n_f$=0) and unquenched ($n_f$=3) 
$\alpha_s$ figures differ markedly and also run differently, 
both in agreement with the perturbative expectation (Mason \etal 2005). 

\subsection{Results on matrix elements}

A key point where lattice QCD calculations are needed and 
can make an impact 
is in the determination of the elements of the 
Cabibbo-Kobayashi-Maskawa matrix that links the quark
flavours under weak decay in the Standard Model of particle 
physics. When a quark changes flavour 
inside a hadron with the emission of a $W$ the quark 
level process is quite simple (see \Figure{\ref{davies:fig24}}). 
However, this 
decay unavoidably takes places inside a hadron because the quarks 
are confined by QCD and the QCD corrections to the decay rate
are significant. This is why the decay matrix element need 
to be calculated in lattice QCD so that all of the gluonic 
radiation around the decay vertex can be taken into account.  
A CKM element $V_{f_1f_2}$ multiplies the vertex but this 
appears in a simple way in the final theoretical answer for the 
decay rate since there is only one weak vertex in the process. 
A comparison of the experimental decay rate and the 
lattice QCD results times $V^2_{f_1f_2}$ then gives $V_{f_1f_2}$. 

Decay rates which can be accurately calculated in lattice 
QCD are those for gold-plated 
hadrons in which there is at most one 
(gold-plated) hadron in the final state. This therefore 
includes leptonic and semi-leptonic decays and the 
mixing of neutral $B$ and $K$ mesons.  Luckily there is a
gold-plated decay mode available to extract each element 
(except $V_{tb}$)
of the CKM matrix which mixes quark flavours under the 
weak interactions in the Standard Model:
\[ \left( \begin{array}{ccc}
{\bf V_{ud}}  & {\bf V_{us}} & {\bf V_{ub}} \\
\pi \rightarrow l\nu & K \rightarrow l \nu & B \rightarrow \pi l \nu \\
 & K \rightarrow \pi l \nu &  \\
{\bf V_{cd}} & {\bf V_{cs}} & {\bf V_{cb}} \\
D \rightarrow l\nu & D_s \rightarrow l \nu & B \rightarrow D l \nu \\
D \rightarrow \pi l\nu & D \rightarrow K l \nu &  \\
{\bf V_{td}} & {\bf V_{ts}} & {\bf V_{tb}} \\
\langle  B_d | \overline{B}_d \rangle  & \langle B_s | \overline{B}_s \rangle  & \\
\end{array} \right) \]
The determination of the CKM elements and 
tests of the self-consistency of the CKM matrix 
 are the current focus for the search for Beyond the Standard 
Model physics and lattice calculations of these decay rates 
will be a key factor in the precision with which this can be done.  

\begin{figure}[ht]
\centering{
\includegraphics[width=90mm, clip, trim=0 0 0 0]{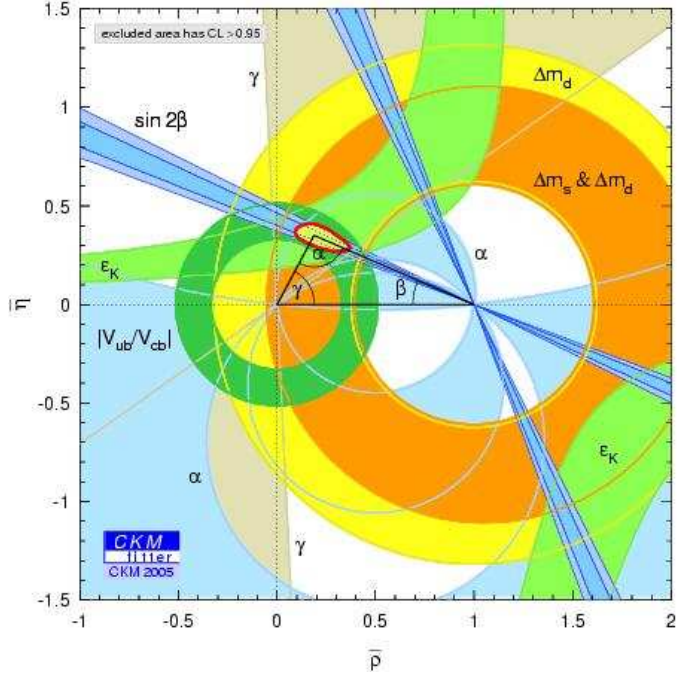}
\caption{Current constraints on the vertex of the `unitarity triangle' 
made from the Cabibbo-Kobayashi-Maskawa matrix. (CKMfitter 2005)}
\label{davies:fig21}}
\end{figure}

\Figure{\ref{davies:fig21}} shows the current `unitarity triangle' picture 
in which limits are placed on various combinations of elements 
of the CKM matrix and the result is expressed as a search 
for the vertex of a triangle. The limits that depend on 
results from $B$ factories and the associated lattice calculations, 
are the dark green ring and the orange and yellow rings. The light 
green hyperbola comes from kaon physics and associated lattice 
calculations. The dark 
green ring is fixed from semileptonic decays of $B$ mesons to 
$\pi$ mesons or $D$ mesons. The orange and yellow rings 
result from mixing of neutral $B$ or $B_s$ mesons. We will 
discuss further below the lattice results for these matrix 
elements. The angles 
of the unitarity triangle, such as $\sin(2\beta)$, are determined directly 
by the experiment (light blue lines) without theoretical input. 

Decay matrix elements of this kind can be calculated on the lattice 
from the amplitudes of the exponentials in the fit functions 
for hadron masses, Equation~\ref{multifit}. For example, the 
rate at which a $B$ meson decays completely to leptons via a 
$W$ boson depends on the matrix element of the heavy-light 
axial vector current, $J_{A_{\mu}} = \overline{\psi}_b\gamma_{\mu}\gamma_5\psi_u$, 
between a $B$ meson and the (QCD) vacuum. The matrix element 
of this current is parameterised by the decay constant, $f_B$, so 
that 
\begin{equation}
\langle 0|J_{A_{\mu}}|B\rangle = p_{\mu}f_B
\end{equation} 
(using a relativistic normalisation of the state).
For a $B$ meson at rest only $J_{A_0}$ is relevant and the 
right-hand-side of the equation above becomes $m_Bf_B$. 

\begin{figure}[ht]
\centering{
\includegraphics[height=27mm, clip, trim=0 0 0 0]{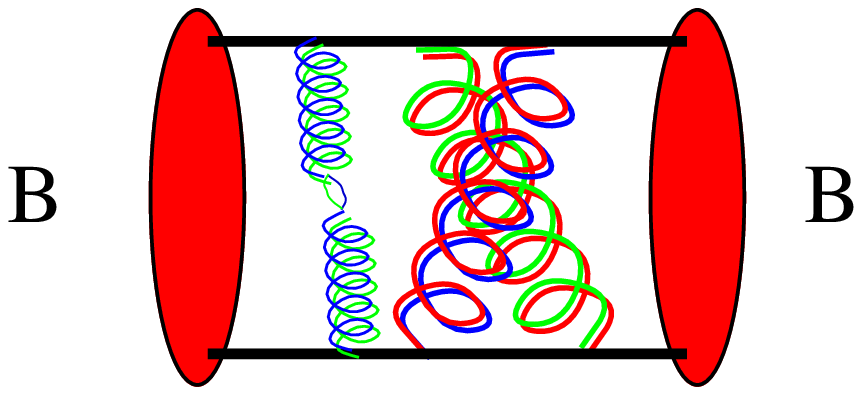}
\includegraphics[height=24mm, clip, trim=0 0 0 0]{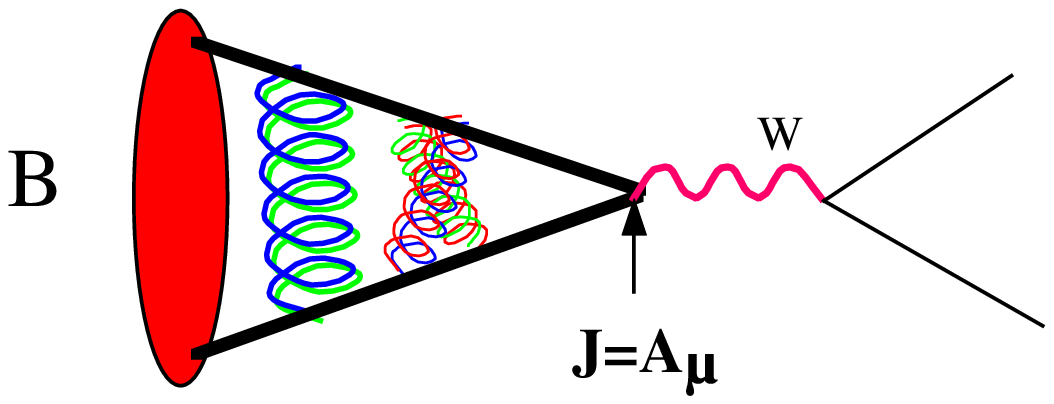}
\caption{(Left) The calculation on the lattice of the 2-point correlator for 
a $B$ meson. 
(Right) The calculation on the lattice of a 2-point function in which 
a $B$ meson decays leptonically.}
\label{davies:fig22}}
\end{figure}

The way in which $f_B$ is calculated on the lattice is illustrated in 
\Figure{\ref{davies:fig22}}. On the left-hand side is an illustration of 
the usual operator (correlator) whose expectation value we 
calculate to determine the $B$ meson mass or energy. 
Then, as earlier, 
\begin{equation}
 << H^{\dag}(T)H(0) >> = \sum_n C_n e^{-E_naT}.
\label{multifit2}
\end{equation}
For the $B$ meson $C_0$ in the fit above is equal to $(\langle 0 | H | B \rangle)^2/2E_0$, 
with a relativistic normalisation of the states.  
On the right-hand-side of \Figure{\ref{davies:fig22}} we illustrate the operator
in which the $B$ meson is destroyed by the axial vector current (the 
$W$ decay to leptons is handled analytically and separately from the 
lattice QCD calculation). Then 
\begin{equation}
\frac{1}{Z}\langle 0 | H^{\dag}(T)J_{A_{\mu}}(0) | 0 \rangle = << H^{\dag}(T)J_{A_{\mu}}(0) >> 
= \sum_n D_n e^{-E_naT}.
\label{current}
\end{equation}
We again have a product of $M^{-1}$ factors to average over configurations, 
since $J_{A_{\mu}}$ contains the same kind of product of $\overline{\psi}$ and 
$\psi$ fields as $H$.
The same energies appear in this fit (indeed the correlators should be fit simultaneously 
to ensure that this is true) but the amplitudes are different. 
Now $D_0 = (\langle 0 | H | B \rangle)(\langle B | J_{A_{\mu}} | 0 \rangle)/2E_0$. 
The second factor is the matrix element that we want and we can extract this 
as $2E_0D_0/\sqrt{(2E_0C_0)}$. If the temporal axial current is used for 
a $B$ meson at rest then $f_B\sqrt{m_B}a^{3/2} = \sqrt{2/C_0}D_0$. 

The current $J_{A_{\mu}}$ that we use on the lattice needs to be well-matched to 
the continuum current. The leading term in the lattice version of 
the current will be just be the obvious transcription of 
the continuum current,
$A_{\mu} = \overline{\psi}_b\gamma_{\mu}\gamma_5\psi_u$. However this 
current has discretisation errors at $\cal{O}$$(\alpha_sa)$ and these 
can be improved by adding higher order operators to cancel the errors 
in the same way as for the action earlier. Again we can use 
perturbation theory to calculate the coefficients of these correction 
terms. If we use NRQCD for the $b$ 
quark in the current then there are also relativistic corrections that 
can be applied to make the current a more accurate version of the 
continuum current. In fact the relativistic correction operators and 
the discretisation correction operators are the same and simply 
pick up perturbative coefficients that are functions of 
$m_Qa$ (Morningstar and Shigemitsu 1998).

We also have to renormalise the matrix elements from the 
lattice to an appropriate continuum renormalisation scheme such 
as $\overline{MS}$. This can be done in perturbation theory 
as, again, it takes account of gluons with momenta above
the lattice cut-off. Such calculations have only been done through 
$\cal{O}$$(\alpha_s)$ so far and this means that the final 
quoted result has errors at $\cal{O}$$(\alpha_s)^2$ (Morningstar and Shigemitsu 1998). Now that 
the systematic error from working in the quenched approximation 
has been overcome, this is often the largest source of error 
and much more work must be done in future to reduce this error. 
Methods for renormalisation and matching that use direct 
numerical methods on the lattice (often called nonperturbative)
are also being explored by many people. 
Matrix elements of conserved currents do not need renormalisation and 
this explains why, for example, the calculations of $f_{\pi}$ 
and $f_K$ described above are so accurate. 

\begin{figure}[ht]
\centering{
\includegraphics[width=90mm, clip, trim=0 0 0 0]{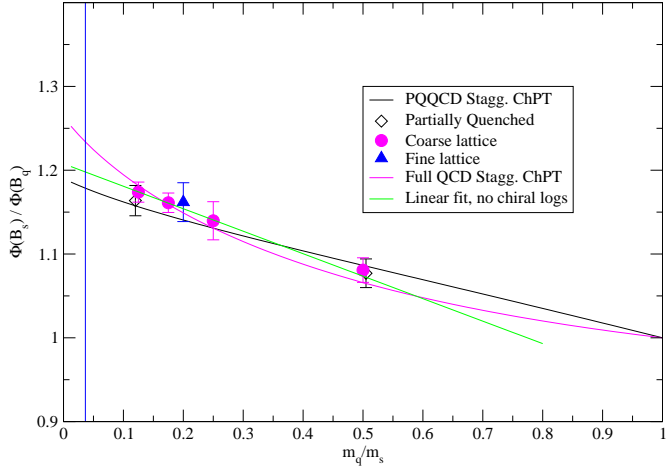}
\caption{The ratio of $f_{B_S}\sqrt{m_{B_s}}/f_B\sqrt{m_B}$ from 
unquenched lattice calculations on the MILC configurations 
at two values of the lattice 
spacing (Gray \etal 2005). The lines represent fits using 
chiral perturbation theory to various combinations of valence 
and sea light quark masses. The pink curve is the full QCD 
curve that extrapolates to the physical answer.}
\label{davies:fig23}}
\end{figure}

The $B$ meson decay constant is of interest, both because it 
sets the rate of $B$ leptonic decay but also because 
it appears in the mixing rate of neutral $B$ mesons. The 
mixing rate is parameterised 
by $f_B^2B_B$ where $B_B$ is the `bag constant'. $f_B^2B_B$ 
is being calculated in lattice QCD but it is harder than the 
calculation of $f_B$ alone. We believe that the $B_B$ factor 
is fairly benign with a value around 1. The calculation of 
$f_B$ then provides a good indicator of the size of mixing 
effects (for the orange and yellow rings in \Figure{\ref{davies:fig21}})
until $f_B^2B_B$ is known accurately.
 
\begin{figure}[ht]
\centering{
\includegraphics[width=65mm, clip, trim=0 0 0 0]{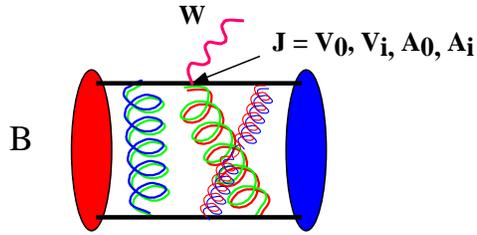}
\caption{The calculation of a 3-point function for $B \rightarrow \pi$ 
semileptonic decay on the lattice.}
\label{davies:fig24}}
\end{figure}

\Figure{\ref{davies:fig23}} shows 
the current results by the HPQCD collaboration, using the 
MILC unquenched configurations and the NRQCD formalism for 
the valence $b$ quarks and asqtad improved staggered formalism 
for the valence light quarks (Gray \etal 2005, Wingate 2004). What is plotted is the ratio of 
$f_{B_s}\sqrt{m_{B_s}}/f_{B}\sqrt{m_B}$ in which the overall 
renormalisation constant for the lattice $J_{A_0}$ cancels giving 
an accurate result. What is interesting here is the approach 
to the light quark mass limit in which $m_{u/d}$ takes its 
physical value. Chiral perturbation theory expects fairly 
strong logarithmic dependence but earlier results worked at 
such heavy $m_{u/d}$ that they were not in the region in 
which chiral perturbation theory was valid. It is clear that 
the new results now have light $m_{u/d}$ and an accurate 
result for this ratio will be possible. 

\begin{figure}[ht]
\centering{
\includegraphics[width=100mm, clip, trim=0 0 0 20]{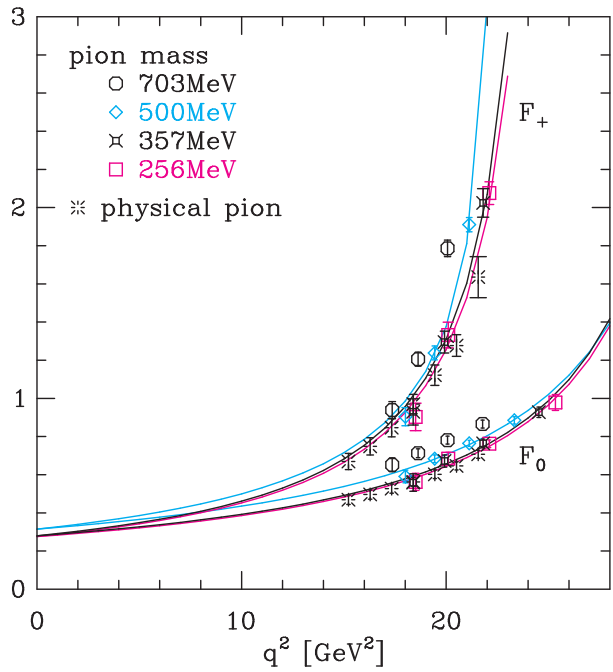}
\caption{Results for the form factors for $B \rightarrow \pi$ 
semileptonic decay from unquenched lattice QCD (Shigemitsu \etal 2005). 
Results are shown for a variety of light quarks, given in terms of 
the mass of the $\pi$ made from these quarks. The bursts show the 
(small) extrapolation to the real $\pi$ mass.}
\label{davies:fig25a}}
\end{figure}

The determination of semileptonic decay rates requires
the calculation of a 3-point function on the lattice. 
This is illustrated in \Figure{\ref{davies:fig24}} for $B \rightarrow \pi$. We now have 
two hadron operators at 0 and $T$ for different hadrons and 
an intermediate current operator $J_{V_{\mu}}$ or 
$J_{A_{\mu}}$ which causes the quark flavour change and 
the emission of a $W$. Now
\begin{equation}
\frac{1}{Z}\langle 0 | H^{\dag}(T)J(t)H^{\prime}(0) | 0 \rangle = << H^{\dag}(T)J(t)H^{\prime}(0) >> 
= \sum_n D_n e^{-E_na(T-t)}e^{-E^{\prime}at}.
\label{3ptcurrent}
\end{equation}
The amplitudes, $D_n$, are now related to $<0|H|B><B|J|\pi><\pi|H^{\prime}|0>$ and 
the matrix element that we want, {\textit{ie}} $<B|J|\pi>$ can be extracted by 
simultaneously fitting the relevant 2-point functions to determine the 
other factors in $D_0$. The matrix element is now a function of the 
4-momentum transfer between $B$ and $\pi$, $q^2$. It is expressed in terms of 
two different form factors, $f_{+}(q^2)$ and $f_0(q^2)$, with different 
momentum-dependent prefactors. $f_{+}(q^2)$ is the form factor that translates
directly into the rate of semileptonic decay, since $f_0$ ends up in this rate 
multiplied by the mass of the lepton into which the $W$ decays, which is very small.  

\begin{figure}[ht]
\centering{
\includegraphics[width=75mm, clip, trim=0 10 0 10]{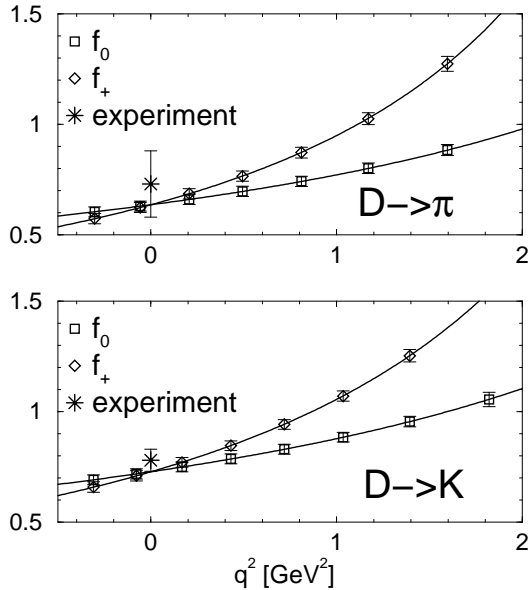}
\caption{Results for the form factors for $D$ semileptonic decay to 
$\pi$ and $K$. 
The experimental points shown at $q^2=0$ 
use the experimental decay rate at that point and the 
current PDG results for the appropriate CKM elements (Aubin \etal 2005).} 
\label{davies:fig25b}}
\end{figure}

Once again we must match the lattice current to a continuum current and 
renormalise to obtain results in a continuum renormalisation scheme. 
This limits the accuracy with which these calculations can now be done 
and more work is required to improve this situation. 
\Figure{\ref{davies:fig25a}} and \Figure{\ref{davies:fig25b}} show current results obtained by the 
HPQCD and FNAL/MILC collaborations for the formfactors for 
$B \rightarrow \pi$ decay and $D \rightarrow \pi/K$ decay respectively. 
The $B$ results use NRQCD $b$ quarks (Shigemitsu \etal 2005); the $D$ results use the 
Fermilab formalism for the $c$ quark (Aubin \etal 2005). The usefulness of the $D$ results 
is to compare to imminent experimental results from the CLEO-c 
collaboration that will provide a check of lattice methods and 
systematic errors for confidence in our precision $B$ results (Shipsey 2005). 

\section{Conclusions}

Lattice QCD has come a long way from the original calculations of 
the 1970s. The original idea that we could solve a simple discretisation 
of QCD numerically
by `brute force' has been replaced by a more sophisticated approach. 
Improved discretisation of both the gluon action and the quark action 
has led to the possibility of performing realistic simulations 
of QCD on current computers. Indeed this has been done, and I hope that 
I have conveyed something of the excitement of seeing accurate 
lattice calculations reproduce well-known experimental numbers 
for the first time. These first calculations are of the masses of 
`gold-plated' hadrons, those for which lattice QCD must be able 
to get the right answer if it is to be trusted at all. Leading 
on from this we have been able to make the first accurate lattice QCD prediction
of the mass of a new meson, 
the $B_c$. 

It is now important to beat down the sources of systematic 
error in the lattice calculation of decay matrix elements for 
$B$ and $D$ physics to obtain results that 
can be combined with experiment to give an accurate 
determination of elements of the CKM matrix. 
The timescale for this programme is the next two years and, as 
well as lattice QCD calculations, it requires $\cal{O}$$(\alpha_s^2)$ 
calculations in perturbation theory to renormalise the 
lattice results to numbers appropriate to the continuous 
real world. 
On a longer timescale (say five years) studies of more complicated baryonic 
matrix elements will be undertaken and 
progress will be made in understanding to what extent 
accurate lattice calculations can be done of some of the more interesting, 
but not gold-plated, particles in the hadron spectrum. 

\section*{Acknowledgements}
It was a pleasure to contribute to this interesting school. 
I am grateful to a large number of lattice colleagues, but 
particularly my long-standing collaborators Peter Lepage and Junko Shigemitsu, 
for numerous useful discussions over many years. 

\section*{References}
There are a number of books on lattice QCD that provide more information 
on the theoretical background, such as Smit, 2002, {\it An introduction 
to Quantum Fields on a lattice}, Cambridge University Press. 
The annual lattice conference provides up-to-date review talks and 
access to the literature. The 2004 Conference Proceedings is published 
in {\textit{Nucl Phys B Proc Suppl}} \vol140. Below I provide a few references, 
concentrating on other summer school lectures where possible. 
\vspace{1cm}

\frenchspacing
\begin{small}

\reference{Acosta D \etal [CDF collaboration], 2005, hep-ex/0505076.}
\reference{Alford M \etal, 1995, \pl {\textit{B}} \vol361 87, hep-lat/9507010.}
\reference{Allison I \etal [HPQCD/FNAL collaborations], 2005, \prl \vol94 172001, hep-lat/0411027.}
\reference{Arndt D, PhD thesis, University of Washington, USA, hep-lat/0406011.}
\reference{Aubin C \etal [MILC collaboration], 2004a, \prd \vol70 094505, hep-lat/0402030.}
\reference{Aubin C \etal [MILC collaboration], 2004b, \prd \vol70 114501, hep-lat/0407028.}
\reference{Aubin C \etal [HPQCD/MILC collaborations], 2004c, \prd \vol70 031504, hep-lat/0405022.}
\reference{Aubin C \etal [FNAL/MILC/HPQCD collaborations], 2005, \prl \vol94 011601,\\ hep-lat/0408306.}
\reference{Bali G, 2000, {\textit{Phys Rep}} \vol343 1, hep-lat/0001312.}
\reference{Bernard C [MILC collaboration], 2001, \prd \vol64 054506, hep-lat/0104002.}
\reference{Chiu T, 2004, {\textit{Nuc Phys B Proc Suppl}} \vol129 135, hep-lat/0310043.}
\reference{CKMfitter, 2005, http://ckmfitter.in2p3.fr/}
\reference{Davies C, 1998, Springer lecture notes in physics, Eds Gausterer, Lang, hep-ph/9710394.}
\reference{Davies C \etal, 2005, {\textit{Nuc Phys B Proc Suppl}} \vol140 261, hep-lat/0409039.}
\reference{Davies C \etal [HPQCD/UKQCD/MILC/FNAL collaborations], 2004, \prl \vol92 022001, hep-lat/0304004.}
\reference{di Pierro  M, 2000, lectures given at the GSA summer school on Physics on the 
Frontier, hep-lat/0009001. Includes code for example calculations. }
\reference{di Pierro M \etal [Fermilab collaboration], 2004, {\textit{Nucl Phys B Proc Suppl}} \vol129 328, hep-lat/0310045;\\
{\textit{Nucl Phys B Proc Suppl}} \vol129 340, hep-lat/0310042.}
\reference{El-Khadra A \etal, 1997, \prd \vol55 3933, hep-lat/9604004.}
\reference{Follana E \etal [HPQCD/UKQCD collaborations], 2004, \prl \vol93 241601,\\ hep-lat/0406010.}
\reference{Frezzotti R, 2005, {\textit{Nuc Phys B Proc Suppl}} \vol140 134, hep-lat/0409138.}
\reference{Gray A \etal [HPQCD/UKQCD collaborations], 2003, {\textit{Nuc Phys B Proc Suppl}} \vol119 592;
\\hep-lat/0507013, accepted for publication in \prd.}
\reference{Gray A \etal [HPQCD collaboration], 2005, {\textit{Nuc Phys B Proc Suppl}} \vol140 446; 
and in preparation.}
\reference{Hein J \etal [HPQCD/UKQCD collaborations], 2003, {\textit{Nucl Phys B Proc Suppl}} \vol119 317, hep-lat/0209077.}
\reference{Ishikawa K \etal [JLQCD/CP-PACS collaborations], 2005, {\textit{Nuc Phys B Proc Suppl}} \vol140 225, hep-lat/0409124.}
\reference{Lepage P, 1996, Schladming winter school, hep-lat/9607076.}
\reference{Lepage P, 1998a. For a useful discussion of path integrals in quantum 
mechanics and their relevance to lattice QCD see \textit{Lattice QCD for novices}, HUGS98, 
hep-lat/0506036. Includes code for example calculations.}
\reference{Lepage P, 1998b, \prd \vol59 074502, hep-lat/9809157.}
\reference{Lepage P \etal, 1992, \prd \vol46, 4052, hep-lat/9205007.}
\reference{Lepage P and Mackenzie P, 1993, \prd \vol48, 2250, hep-lat/9209022.}
\reference{Lepage P and Davies C, 2004, {\textit{Int J Mod Phys A}} \vol19 877.}
\reference{Mason Q \etal [HPQCD/UKQCD collaborations], 2005, \prl \vol95 052002,\\ hep-lat/0503005.}
\reference{Morningstar C and Shigemitsu J, 1998, \prd \vol57 6741, hep-lat/9712016.}
\reference{Nobes M, 2005, hep-lat/0501009.}
\reference{Nobes M \etal, 2005, in preparation.}
\reference{Orginos K \etal, 1999, \prd \vol60 054503, hep-lat/9903032.}
\reference{PDG, 2004, http://pdg.lbl.gov/}
\reference{Shigemitsu J \etal [HPQCD collaboration], 2005, {\textit{Nuc Phys B Proc Suppl}} \vol140 464,\\ hep-lat/0408019.}
\reference{Shipsey I, 2005, {\textit{Nuc Phys B Proc Suppl}} \vol140 58, hep-lat/0411009.}
\reference{Sommer R, 1994, {\textit{{Nuc Phys B}} \vol411 839, hep-lat/9310022.}
\reference{Toussaint D and Davies C, 2005, \textit{Nuc Phys B Proc Suppl}} \vol140 234, hep-lat/0409129.}
\reference{Trottier H, 2004, {\textit{Nuc Phys B Proc Suppl}} \vol129 142, hep-lat/0310044.}
\reference{Wingate M [HPQCD collaboration], 2004, \prl \vol92 162001, hep-lat/0311130.}
\end{small}
\nonfrenchspacing
\end{document}